\documentclass[12pt]{article}
\usepackage{latexsym}
\usepackage{amsmath}
\usepackage{amssymb}
\usepackage{relsize}
\usepackage{geometry}

\numberwithin{equation}{section}

\usepackage{tensor}
\usepackage{physics} 
\usepackage{csquotes} 

\usepackage{graphicx}       	
\graphicspath{{./img/}} 
\usepackage[font=small,labelfont=bf]{caption}
\usepackage{subcaption}

\def\beqn{\begin{eqnarray}}
\def\eeqn{\end{eqnarray}}

\def\beq{\begin{equation}}
\def\eeq{\end{equation}}
\def\ba{\beq\new\begin{array}{c}}
\def\ea{\end{array}\eeq}

\newcommand{\gsim}{\lower.7ex\hbox{$
\;\stackrel{\textstyle>}{\sim}\;$}}
\newcommand{\lsim}{\lower.7ex\hbox{$
\;\stackrel{\textstyle<}{\sim}\;$}}

\newcommand{\ntwo}{${\mathcal N}=2$ }

\newcommand{\ntwot}{${\mathcal N}= \left(2,2\right) $ }

\newcommand{\pt}{\partial}

\def\slashed#1{\setbox0=\hbox{$#1$}             
   \dimen0=\wd0                                 
   \setbox1=\hbox{/} \dimen1=\wd1               
   \ifdim\dimen0>\dimen1                        
      \rlap{\hbox to \dimen0{\hfil/\hfil}}      
      #1                                        
   \else                                        
      \rlap{\hbox to \dimen1{\hfil$#1$\hfil}}   
      /                                         
   \fi}                                        %

\newcommand{\wcpt}{$\mathbb{WCP}(2,2)\;$}
\newcommand{\wcpo}{$\mathbb{WCP}(1,1)\;$}

\usepackage{mathtools}
\usepackage{hyperref}
\usepackage{xcolor}
\usepackage{tikz-cd}
\usepackage{placeins}
\usepackage[toc,page, title, titletoc]{appendix}
\newcommand{\mathsym}[1]{{}}
\newcommand{\unicode}[1]{{}}

\begin{document}

\hypersetup{%
linkbordercolor=blue,
}

%
%

\begin{titlepage}

\begin{flushright}
FTPI-MINN-24-33 \\ UMN-TH-4415/24
\end{flushright} 

\begin{center}

{\Large{\bf 
Search for NS 3-form Flux Induced Vacua for the Critical Non-Abelian Vortex String
}}

\vspace{5mm}

{\large  \bf E.~Ievlev$^{\,a}$, P.~Pichugina$^{\,b,c}$ and  A.~Yung$^{\,b,d}$}
\end{center}

\begin{center}

{\it  $^{a}$William I. Fine Theoretical Physics Institute,
University of Minnesota,
Minneapolis, MN 55455, USA}\\

$^{b}${\it National Research Center ``Kurchatov Institute'',
Petersburg Nuclear Physics Institute, Gatchina, St. Petersburg
188300, Russia}\\

$^{c}${\it  St. Petersburg State University, Universitetskaya nab., St.~Petersburg \\ 199034, Russia}\\

{\it $^d$Higher School of Economics, National Research University, St. Petersburg \\
194100, Russia}

\end{center}

\vspace{5mm}

\begin{center}
{\large\bf Abstract}
\end{center}

It has been demonstrated that the non-Abelian solitonic vortex string in four-dimensional (4D) $\mathcal{N}=2$ supersymmetric QCD (SQCD) with gauge group U(2) and $N_f=4$ quark hypermultiplets behaves as a critical superstring. This string propagates in a ten-dimensional space comprising the flat 4D space and an internal Calabi-Yau noncompact threefold, specifically, the conifold. The lowest state of this string is a massless BPS baryon associated with the deformation of the conifold's complex structure modulus, $b$. Previous studies considered deformations of the 10-dimensional background by a nonzero Neveu-Schwarz (NS) 3-form flux, which was interpreted as selecting specific quark masses in 4D SQCD. These deformations and the corresponding back reaction on the metric were analyzed at the leading order at small 3-form flux. In this paper, we first derive the exact potential for the conifold complex structure modulus $b$ generated by the 3-form flux assuming a fixed conifold background. Then we include the back reaction effects and solve the corresponding gravity equations of motion. We show that the resulting potential gives a runaway vacuum. At this runaway vacuum the conifold undergoes degeneration.


\end{titlepage}

\setcounter{tocdepth}{2}
\numberwithin{equation}{section}
\tableofcontents

\clearpage

\section {Introduction }
\label{intro}

Non-Abelian vortex strings, first found in \cite{HT1,ABEKY,SYmon,HT2}, are the confining strings of the 4D \ntwo SQCD with the gauge group U$(N)$ and $N_f \ge N$ hypermultiplet quark  flavors.
These strings are 1/2 Bogomolny-Prasad-Sommerfeld (BPS) saturated and, therefore, possess \ntwot supersymmetry on the world sheet.
Besides the usual translational moduli of the Abrikosov-Nielsen-Olesen (ANO)-type strings \cite{ANO}, they also carry additional moduli corresponding to orientation of the color flux and the size of the string \cite{HT1,ABEKY,SYmon,HT2} (see \cite{Trev,Jrev,SYrev,Trev2} for reviews). 

In \cite{SYcstring} it was discovered that the non-Abelian vortex string becomes critical in the theory with $N=2$ colors and $N_f=4$ quark flavors.
In this case, the 4D SQCD $\beta$-function vanishes to all orders in perturbation theory, and no dynamical scale $\Lambda$ is generated.
The theory living on the string world sheet in this case is a product of a free translational sector and a weighted $\mathbb{CP}$ model ($\mathbb{WCP}(2,2)$ for short) for the internal moduli, see below.
Its $\beta$-function also vanishes, and the overall Virasoro central charge is critical \cite{SYcstring}.  
This happens because in addition to four translational moduli, non-Abelian string has six orientational and size moduli. 
Together, they form a ten-dimensional target space required for a superstring to be critical. 
The target space of the string sigma model is $\mathbb{R}_4\times Y_6$, a product of the flat four-dimensional space and a Calabi-Yau (CY) non-compact threefold $Y_6$, namely, the conifold \cite{Candel,NVafa}. Moreover, the theory of the critical vortex string at hand was identified as the superstring theory of type IIA \cite{KSYconifold}. 

This allows one to apply string theory for consideration of the closed string spectrum and to interpret it  as a spectrum of hadrons in 4D \ntwo SQCD.

The study of the above vortex string from the standpoint of string theory, with the focus on massless states in four dimensions has been started in \cite{KSYconifold,KSYcstring} using supergravity approximation.
It turns out that  most of massless modes have  non-normalizable wave functions over the non-compact conifold $Y_6$, i.e. they are not localized in 4D and, hence, cannot be interpreted as dynamical states in 4D SQCD. 
In particular, the 4D graviton and  unwanted vector multiplet associated with deformations of the K\"ahler form of the conifold are absent.
 However, a single massless BPS hypermultiplet   was found 
at the  self-dual point at strong coupling. It is associated with deformations of a complex structure of the conifold and was  interpreted  as a composite 4D baryon $b$ \footnote{ The definition of the baryonic charge is non-standard and will be given below in Sec. \ref{sec:NAstring}.}.
The vacuum expectation value (VEV) of $b$ parametrizes a new non-perturbative Higgs branch of the 4D SQCD arising at strong coupling.
Later  low lying massive non-BPS 4D states were found in \cite{SYlittles,SYlittmult} using the little string theory 
approach, see \cite{Kutasov} for a review.

In previous works \cite{Y_NSflux,NS} a study of possible flux deformations of the 10D background for non-Abelian vortex string was initiated. 
The goal is to look for flux deformations of the string background which do not destroy \ntwo supersymmetry in 4D and interpret them in terms of  certain deformations in SQCD. Fluxes generically induce a potential for CY moduli lifting flat directions, see, for example, \cite{Louis} for a review. It is known that for type IIA CY compactifications the potential for the  K\"ahler form moduli arise from Ramond-Ramond (RR) even-form fluxes,
while  the potential for complex structure moduli  is induced by the Neveu-Schwarz (NS) 3-form flux $H_3$  \cite{Louis2,Kachru}.
Since for the conifold case at hand the only modulus associated with a physical state is the complex structure modulus $b$ we focus on the NS 3-form flux. It does not break \ntwo supersymmetry in 4D theory \cite{Louis2}.

In \cite{Y_NSflux} the NS 3-form flux $H_3$ was interpreted as switching on quark masses in 4D SQCD.
The motivation is that the only scalar potential deformation, which is allowed in SQCD by \ntwo supersymmetry is the mass term for quarks. 
Field theory arguments were used to find a particular choice of nonzero quark masses associated with $H_3$.

The flux deformation was studied  in \cite{Y_NSflux} to the leading order at small $H_3$ which translates into small values of quark masses.
A subsequent work \cite{NS} studied the back reaction of the nonzero 3-form flux on the metric and dilaton. 
An ansatz with several warp factors was introduced.
The linearized gravity equations of motion were solved in an asymptotic region far from the conifold apex, which allowed to compute the potential generated by the 3-form flux for the conifold complex structure modulus  $b$ in the regime when  $b$ is large enough.
It was found that the resulting potential for $b$ leads to a runaway vacuum. 
At the runaway vacuum warp factors disappear, while the deformed conifold degenerates. In 4D SQCD this was interpreted as a   flow to U(1) gauge theory upon switching on quark masses and decoupling of two flavors. 

In the first part of this paper we revisit the computation of the potential to the leading order at small $H_3$.
Previously in \cite{Y_NSflux} solutions for the 3-form on the deformed conifold were found asymptotically in the regions of very small  and very large distances  from the conifold apex.
It was not clear whether these solutions are smoothly connected.
Under the assumption that they do, an approximate form of the potential $V(b)$ was computed.
Here we close that gap and find an exact solution for the 3-form on the deformed conifold background.
We find that the assumption of \cite{Y_NSflux} was correct.
We also compute the exact potential $V(b)$ and find that it agrees with the approximate result of \cite{Y_NSflux} in the limits mentioned above.

In the second part of the current work we study the effects of back reaction of 3-form $H_3$ on the metric and dilaton, relaxing the assumption of large $b$.
We solve full non-linear gravity equations of motion combining analytical and numerical methods.  Using these solutions  we  find the potential $V(b)$ at small and large $b$ looking for all possible vacua compatible with  our ansatz  for the metric.
Our finding is that no new vacua appear and taking   non-linear effects into account do not  change the conclusions of \cite{Y_NSflux,NS}. Namely,  the  $H_3$ flux generates a potential leading to  the runaway vacuum for the baryonic scalar field  $b$.

The paper is organized follows. 
In Sec.~\ref{sec:NAstring} we review the construction of critical non-Abelian strings in 4D \ntwo SQCD.
In Sec.~\ref{sec:no_backreaction} we review the low energy supergravity setup, find an exact solution for the 3-form $H_3$ on the deformed conifold  and derive an exact baryon potential assuming a fixed conifold background.
After that we include back reaction effects in Sec.~\ref{sec:backreact}.
In Sec.~\eqref{sec:quarkmasses} we interpret our results in terms of 4D SQCD.
Sec.~\ref{sec:concl} presents our conclusions.

\section {Non-Abelian critical vortex string}
\label{sec:NAstring}

In this Section we briefly review the non-Abelian string construction and the non-perturbative Higgs branch of 4D SQCD.

\subsection{Four-dimensional ${\mathcal N}=2$ SQCD}
\label{sec:SQCD}

As was mentioned in the Introduction, the non-Abelian vortex string was initially identified in 4D
\ntwo SQCD with a U$(N)$ gauge group and $N_f \ge N$ quark flavors incorporating a Fayet-Iliopoulos (FI)  term  \cite{FI} with parameter $\xi$
\cite{HT1,ABEKY,SYmon,HT2}, see e.g. \cite{SYrev} for a comprehensive review.
In brief, at weak coupling ($g^2 \ll 1$), the theory enters a Higgs phase where the scalar components of the quark multiplets (squarks) acquire vacuum expectation values (VEVs), leading to the breaking of the U$(N)$ gauge group. This Higgs mechanism gives mass to all gauge bosons, which combine with screened quarks to form extended \ntwo multiplets with mass $m_G \sim g\sqrt{\xi}$.

The global flavor symmetry SU$(N_f)$ is reduced to the so-called color-flavor locked group. Consequently, the global symmetry of the system is 
\beq
 {\rm SU}(N)_{C+F}\times {\rm SU}(N_f-N)\times {\rm U}(1)_B,
\label{c+f}
\eeq
The unbroken global U(1)$_B$ factor mentioned above corresponds to a baryonic symmetry. 
Typically, what is recognized as the baryonic U(1) charge forms part of the gauge group in our 4D theory. 
Here, however, ``our'' U(1)$_B$ arises as an unbroken combination of two U(1) symmetries: the first is a subgroup of the flavor SU$(N_f)$, and the second is the global U(1) subgroup of the U$(N)$ gauge symmetry.

As noted, we are considering \ntwo SQCD in the Higgs phase, where $N$ squarks condense, leading to the confinement of monopoles by non-Abelian vortex strings. 
In this 4D \ntwo theory, these strings are 1/2 BPS-saturated, with their tension exactly given by the FI parameter:
\beq
T=2\pi \xi\,.
\label{ten}
\eeq
Since non-Abelian strings cannot be broken, monopoles cannot attach to the endpoints of these strings. 
Instead, in U$(N)$ theories, confined monopoles act as junctions between two distinct types of elementary non-Abelian strings \cite{SYmon,HT2,T} (for a detailed review, see \cite{SYrev}).
Consequently, in 4D \ntwo SQCD, we observe monopole-anti-monopole mesons, where the monopole and anti-monopole are linked by two confining strings.

Additionally, the U$(N)$ gauge theory can host baryons in the form of closed "necklace" configurations, consisting of $N \times$ (integer) monopoles \cite{SYrev}. 
For the U(2) gauge group, a massless BPS baryon $b$, identified via string theory in \cite{KSYconifold}, comprises four monopoles \cite{ISY_b_baryon}.

Here, we focus on the specific case of $N=2$ and $N_f=4$, since, as mentioned in the Introduction, in this setup 4D \ntwo SQCD admits non-Abelian vortex strings that exhibit properties of critical superstrings \cite{SYcstring}. 
Additionally, for $N_f=2N$, the gauge coupling $g^2$ in 4D SQCD remains fixed as the $\beta$ function vanishes, meaning the coupling does not run. 
However, the 4D theory's conformal invariance is explicitly broken by the FI parameter $\xi$, which sets the quark VEVs. 
Importantly, this FI parameter is not subject to renormalization.

Both stringy monopole-antimonopole mesons and monopole baryons, with spins $J \sim 1$, have masses proportional to the string tension, scaling as $\sim \sqrt{\xi}$. 
At weak coupling ($g^2 \ll 1$), these masses are larger than those of perturbative states, which have masses on the order of $m_G \sim g \sqrt{\xi}$. 
As a result, these stringy states can decay into lighter perturbative states\footnote{Their quantum numbers with respect to the global symmetry group in \eqref{c+f} permit these decays, as discussed in \cite{SYrev}.}. 
Therefore, at weak coupling, these states are not expected to be stable.

Only in the   strong coupling domain $g^2\sim 1$  we expect that (at least some of) stringy mesons and baryons become stable.
These expectations were confirmed in \cite{KSYconifold,SYlittles} where low lying string states in the string theory for the critical non-Abelian vortex were found at the self-dual point at strong coupling.

Below in this paper we introduce  quark masses $m_A$, $A=1,...4$ assuming that two first squark flavors with masses $m_1$ and $m_2$ develop VEVs.

\subsection{World-sheet sigma model}
\label{sec:wcp}

The color-flavor locked group SU$(N)_{C+F}$ underlies the formation of non-Abelian vortex strings \cite{HT1,ABEKY,SYmon,HT2}. 
A key characteristic of these vortices is the presence of orientational zero modes. 
As mentioned earlier, in \ntwo SQCD, these strings are 1/2 BPS-saturated and preserve \ntwot supersymmetry on their world sheet.

Let us briefly review the model that emerges on the world sheet of the non-Abelian string \cite{SYrev}.

The translational moduli fields of the vortex are governed by the Nambu-Goto action and decouple from the other moduli fields. 
Here, we focus on the internal moduli.

When $N_f = N$, the dynamics of the orientational zero modes of the non-Abelian vortex, which translate into orientational moduli fields on the world sheet, are described by the 2D \ntwot supersymmetric ${\mathbb{CP}}(N-1)$ model.
If additional quark flavors are introduced, the non-Abelian vortices become semilocal and gain size moduli \cite{AchVas}. 
Specifically, for the non-Abelian semilocal vortex in U(2) \ntwo SQCD with four flavors, we have  both the complex orientational moduli $n^P$ (with $P=1,2$) and two complex size moduli $\rho^K$ (with $K=3,4$), see \cite{HT1,HT2,AchVas,SYsem,Jsem,SVY}.

The effective theory on the string world sheet is a two-dimensional \ntwot supersymmetric \wcpt model. For further details, see the review in \cite{SYrev}.
This model can be  defined as a low energy limit of the  U(1) gauge theory \cite{W93}, which corresponds
to taking the limit of infinite gauge coupling, $e^2 \to\infty$, in the action below.
The bosonic part of the action in Euclidean notation reads
\begin{equation}
\begin{aligned}
	&S = \int d^2 x \left\{
	\left|\nabla_{\alpha} n^{P}\right|^2 
	+\left|\tilde{\nabla}_{\alpha} \rho^K\right|^2
	+\frac1{4e^2}F^2_{\alpha\beta} + \frac1{e^2}\,
	\left|\pt_{\alpha}\sigma\right|^2
	\right.
	\\[3mm]
	&+\left.
	2\left|\sigma+\frac{m_P}{\sqrt{2}}\right|^2 \left|n^{P}\right|^2 
	+ 2\left|\sigma+\frac{m_{K}}{\sqrt{2}}\right|^2\left|\rho^K\right|^2
	+ \frac{e^2}{2} \left(|n^{P}|^2-|\rho^K|^2 - {\rm Re}\,\beta \right)^2
	\right\},
	\\[4mm]
	&
	P=1,2\,,\qquad K=3,4\,.
\end{aligned}
\label{wcp22}
\end{equation}
Here $\beta$ is the inverse coupling constant,  $m_A$ ($A=1,..,4$) are the so-called twisted masses (they coincide with  4D quark masses). 
The complex scalar $\sigma$ is a superpartner of the gauge field $A_\alpha$. 
Orientation moduli become complex fields $n^P$, size moduli become fields $\rho^K$  with the $U(1)$ charges $+1$ and $-1$, respectively.
Their covariant derivatives are defined as 
\begin{equation}
	\nabla_{\alpha}=\pt_{\alpha}-iA_{\alpha}\,,
	\qquad 
	\tilde{\nabla}_{\alpha}=\pt_{\alpha}+iA_{\alpha}\,,	
\end{equation}
The  target space of the \wcpt model
is defined by the $D$-term condition
\beq
|n^{P}|^2-|\rho^K|^2 = {\rm Re}\,\beta, \qquad P=1,2, \qquad K=3,4.
\label{D-term}
\eeq
The number of real bosonic degrees of freedom in the model  \wcpt  (the dimension of the Higgs branch) is $8-1-1=6$.  Here 8 is the number of real degrees of 
freedom of $n^P$ and $\rho^K$ fields and we subtracted one real constraint imposed by the the $D$ term condition in 
\eqref{D-term}  and one  gauge phase eaten by the Higgs mechanism. As we already mentioned, these six internal degrees of freedom in the massless limit can be combined with four translational moduli to form a 10D space needed for a superstring to be critical. 

The global symmetry of the world sheet  \wcpt model is
\begin{equation}
	 {\rm SU}(2)\times {\rm SU}(2)\times {\rm U}(1)_B \,,
\label{globgroup}
\end{equation}
i.e. exactly the same as the unbroken global group in the 4D theory at $N=2$ and $N_f=4$. 
The fields $n$ and $\rho$ 
transform in the following representations:
\begin{equation}
	n:\quad \left(\textbf{2},\,\textbf{1},\, \frac12\right), \qquad \rho:\quad \left(\textbf{1},\,\textbf{2},\, \frac12\right)\,.
\label{repsnrho}
\end{equation}
Here  the global ``baryonic''  U(1)$_B$ group   rotates $n$ and 
$\rho$ fields with the same phase,  see \cite{KSYconifold} for details.

Twisted masses of $n^P$ and $\rho^K$ fields coincide with quark masses of 4D SQCD and are given respectively by $m_P$ and 
$m_K$, $P=1,2$ and $K=3,4$, see \cite{SYrev}.
Non-zero twisted masses $m_A$ break each of the SU(2) factors in \eqref{globgroup} down to U(1).

The 2D coupling constant ${\rm Re}\,\beta$ can be naturally complexified to the complex coupling constant $\beta$ if we
include the $\theta$ term in the action \cite{W93}.
At  the quantum level, the coupling $\beta$ does not run in this theory. Thus, 
the \wcpt  model is superconformal at zero masses $m_A = 0$. Therefore, its target space is Ricci flat and  (being K\"ahler due to \ntwot supersymmetry) represents  a non-compact Calabi-Yau manifold,  namely the conifold $Y_6$, see \cite{NVafa} for a review.

The \wcpt model  with $m_A=0$ was used in \cite{SYcstring,KSYconifold} to define 
the critical string theory for the non-Abelian vortex at hand.
Nonzero twisted masses $m_A\neq 0$ define  a mass deformation of the  superconformal  theory
on the conifold. Generically  quark masses break the world sheet conformal invariance. The  \wcpt model with nonzero $m_A$   can no longer be used to define a string theory for the non-Abelian vortex in the massive 4D SQCD.

Therefore, as we already explained in the Introduction, instead
of attempting to interpret twisted mass terms in \eqref{wcp22} in terms of 10D gravity, we take a different route.
Namely, following \cite{Y_NSflux,NS} we look for  solutions of 10D gravity equations of motion with nonzero NS 3-form, 
and then interpret $H_3$ in terms of quark masses. Note that the 3-form $H_3$  flux does not break \ntwo  supersymmetry
in 4D  \cite{Louis2}. Moreover, finding solutions of gravity equations of motion with nonzero flux will garantee that the world sheet theory is conformal in the obtained  background.

\subsection {Massless 4D baryon}
\label{conifold}

In this Section, we briefly review the only 4D massless state identified in the string theory of the critical non-Abelian vortex in the massless limit \cite{KSYconifold}. 
This state is linked to the deformation of the conifold’s complex structure. 

As mentioned earlier, all other massless string modes possess non-normalizable wave functions over the conifold. 
For instance, a 4D graviton associated with a constant wave function over the conifold $Y_6$ is absent \cite{KSYconifold}. 
This result aligns with our expectations, as we began with \ntwo SQCD in a flat, four-dimensional space without gravity.

We can construct the U(1) gauge-invariant ``mesonic'' variables
\beq
w^{PK}= n^P \rho^K.
\label{w}
\eeq
These variables are subject to the constraint
\beq
{\rm det}\, w^{PK} =0. 
\label{coni}
\eeq

Equation (\ref{coni}) defines the conifold $Y_6$.  
It has the K\"ahler Ricci-flat metric and represents a non-compact
 Calabi-Yau manifold \cite{Candel,NVafa,W93}. It is a cone which can be parametrized 
by the non-compact radial coordinate 
\beq
\widetilde{r}^{\, 2} = {\rm Tr}\, \bar{w}w\,
\label{tilder}
\eeq
and five angles, see \cite{Candel}. Its section at fixed $\widetilde{r}$ is $S_2\times S_3$.

At $\beta = 0$, the conifold develops a conical singularity, allowing both spheres, $S_2$ and $S_3$, to shrink to zero size. 
This singularity can be smoothed out in two distinct ways: by deforming the Kähler form or by changing the complex structure.
The first option, called the resolved conifold, involves keeping a non-zero value of $\beta$ in \eqref{D-term}. 
This resolution preserves both the Kähler structure and Ricci-flatness of the metric. 
Setting $\rho^K = 0$ in \eqref{D-term} reduces the model to a $\mathbb{CP}(1)$ model with $S_2$ as the target space (with radius $\sqrt{\beta}$). 

The resolved conifold does not have normalizable zero modes. 
Specifically, the modulus $\beta$, which acts as a scalar field in four dimensions, has a non-normalizable wave function over $Y_6$ and is therefore not dynamical \cite{KSYconifold}.

If $\beta = 0$, there is an alternative option: deforming the complex structure \cite{NVafa}. 
This deformation also preserves the Kähler structure and Ricci-flatness of the conifold and is commonly known as the deformed conifold, defined by modifying Eq.~(\ref{coni}), namely,   
\beq
 {\rm det}\, w^{PK} = b\,,
\label{deformedconi}
\eeq
where $b$ is a complex parameter.
Now  the sphere $S_3$ can not shrink to zero, its minimal size is determined by $b$. 

The modulus $b$ becomes a 4D complex scalar field. The  effective action for  this field was calculated in \cite{KSYconifold}
using the explicit metric on the deformed conifold  \cite{Candel,Ohta,KlebStrass} \footnote{See also \cite{Strom} where this 
logarithmic metric was obtained earlier in a different setup.},
\beq
S_{{\rm kin}}(b) = T\int d^4x |\pt_{\mu} b|^2 \,
\log{\frac{\widetilde{R}_{\rm IR}^2}{|b|}}\,,
\label{Sb}
\eeq
where $\widetilde{R}_{\rm IR}$ is the  maximal value of the radial coordinate $\widetilde{r}$  introduced as an infrared regularization of the 
logarithmically divergent $b$-field  norm. Here the logarithmic integral at small $\widetilde{r}$ is cut off by the minimal size of $S_3$, which is equal to $|b|$.

To clarify, in the standard AdS/CFT lore, the radial coordinate in the internal dimensions represents energy, with large values corresponding to the ultraviolet region. 
In our setup, however, it is the opposite. 
The radial coordinate $\widetilde{r}$ here measures the absolute values of products $n^P \rho^K$. 
Since the $\rho$ fields are vortex string size moduli \cite{AchVas}, $\widetilde{r}$ has a 4D interpretation as the distance from the string axis, with large $\widetilde{r}$ corresponding to the infrared region.

We observe that the norm of the modulus $b$ is logarithmically divergent in the infrared. 
Such modes, with logarithmic divergence, lie on the borderline between normalizable and non-normalizable modes.
Conventionally, such states are regarded as ``localized'' in 4D, and we adopt this interpretation here. 
This scalar mode $b$ is localized near the conifold singularity, similar to how the orientational and size zero modes are localized on the vortex string solution (see \cite{SVY} for more detail).

Being massless, the field $b$ can acquire a vacuum expectation value (VEV), leading to a new Higgs branch in 4D \ntwo SQCD. 
This branch develops only at the critical value of the 4D coupling constant, $\tau_{SW} = 1$, which corresponds to $\beta = 0$.
 From now on we put $\beta=0$ and study the string theory for the non-Abelian critical string on the deformed conifold.

 In \cite{KSYconifold} the massless state $b$ was interpreted as a baryon of 4D \ntwo QCD.
Let us explain this.
 From Eq.~(\ref{deformedconi}) we see that the complex 
parameter $b$ (which is promoted to a 4D scalar field) is a singlet with respect to both SU(2) factors in
 (\ref{globgroup}), i.e. 
the global world-sheet group.\footnote{Which is isomorphic to the 4D
global group \eqref{c+f} for $N=2$, $N_f=4$.} What about its baryonic charge? From \eqref{repsnrho} and \eqref{deformedconi}
we see that the $b$ state transforms as 
\beq
({\bf 1},\,{\bf 1},\,2).
\label{brep}
\eeq
 In particular it has the baryon charge $Q_B(b)=2$.

 In type IIA superstring compactifications the complex scalar 
associated with deformations of the complex structure of the Calabi-Yau
space enters as a 4D \ntwo BPS hypermultiplet, see \cite{Louis} for a review. 

On the field theory side we know that if we switch on generic quark masses in 4D SQCD the $b$-baryon becomes massive. 
Since it is a BPS state its mass is dictated by its baryonic charge \cite{ISY_b_baryon},
\beq
m_{b} = |m_1+m_2-m_3-m_4|.
\label{m_b}
\eeq

\section{Flux  deformation with  a small NS 3-form 
}
\label{sec:no_backreaction}

In this paper we  use the effective 10D supergravity approach to find
the flux-deformed background for our critical non-Abelian vortex
string.  As a first step in this section we consider small  flux and solve equations of motion for $H_3$ using the
conifold metric, neglecting the back reaction of $H_3$ on the metric
and the dilaton. These effects appear in the quadratic order in $H_3$.

We assume that the conifold complex structure
modulus b is large enough to make sure that the curvature
of the conifold is small everywhere. This justifies the gravity approximation.

We start by introducing the supergravity setup and reviewing the asymptotic solutions for the 3-form found in \cite{Y_NSflux}.
After that, we will derive the exact solution for $H_3$ in the fixed background of the deformed conifold  and compute the potential for $b$.

\subsection{Supergravity setup and the singular conifold}
\label{sec:H3_large_r}

The bosonic part of the action of the type IIA supergravity in the Einstein frame is given by
\begin{equation}
	S_{10D} =\frac{1}{2 \kappa ^2} \int d^{10} x \sqrt{-G} \left\{R-\frac{1}{2} G^{M N} \partial_M \Phi \partial_N \Phi- \frac{e^{-\Phi}}{12} H_{MNL} H^{MNL}\right\},
\label{10Daction}
\end{equation}
where $G_{MN}$ and $\Phi$ are the 10D metric and dilaton.
For the constant  dilaton background $\Phi_0$ (at $H_3=0$) the string coupling is given by $g_s=e^{\Phi_0}$. 
The gravitational constant $\kappa^2$ in 10D has mass dimension $-8$; in our conventions, it is related to the tension of the 4d confining string $T$ as
\begin{equation}
	2\kappa^2= \frac{(2\pi)^3 g_s^2}{T^4}
\label{kappa_T_relation}
\end{equation}
Since in this paper we are going to turn on the flux only for the $H_3$ form, in eq.~\eqref{10Daction} we keep (besides the graviton and the dilaton) only the field strength for the corresponding NS 2-form $B_2$, $H_3=dB_2$.

The 10D space has the structure $\mathbb{R}^{1,3} \times Y_6$ with the metric 
\begin{equation}
	ds^2_{10} = -(dt)^2 + (d x^1)^2 + (d x^2)^2 + (d x^3)^2  + \,ds^2_6,
\label{10metric}
\end{equation}
The first piece here is the standard Minkowski flat space metric.
The second piece, $ds^2_6$, is the metric on the singular conifold $Y_6$  given by \cite{Candel}
\begin{equation}
	ds^2_{6}=dr^2 + \frac{r^2}{6}(e_{\theta_1}^2+ e_{\varphi_1}^2 +e_{\theta_2}^2+ e_{\varphi_2}^2) +\frac{r^2}{9}e_{\psi}^2 ,
\label{conmet}
\end{equation}
where
\begin{equation}
\begin{aligned}
	e_{\theta_1} &= d\theta_1 , \qquad  e_{\varphi_1}= \sin{\theta_1}\, d\varphi_1 \,,\\ 
	e_{\theta_2} &= d\theta_2 , \qquad  e_{\varphi_2}= \sin{\theta_2}\, d\varphi_2	\,, \\
	e_{\psi} &= d\psi  + \cos{\theta_1}d\varphi_1+ \cos{\theta_2}d\varphi_2	\,.
\end{aligned}
\label{angles}
\end{equation}
Here $r$ the radial coordinate on the cone, while the angle coordinates are defined at $0\le \theta_{1,2}<\pi$, $0\le \varphi_{1,2}<2\pi$, $0\le \psi<4\pi$. 
As mentioned above, non-zero $b$ will smooth out the conifold singularity and deform the metric (see Sec.~\ref{sec:def} below).
However, far away from the origin, $r\gg |b|^{1/3}$, the metric is still well approximated by the singular conifold metric.
As a warm up, we start our analysis from this region.

The corresponding volume form on the conifold $Y_6$ reads
\begin{equation}
	({\rm Vol})_{Y_6} = \frac{1}{108}\int r^5 \,dr \, d\psi\,  d\theta_1\,\sin{\theta_1} d\varphi_1 \,d\theta_2 \,
	 \sin{\theta_2}d\varphi_2\,.
\label{Vol}
\end{equation}
The radial coordinate $\widetilde{r}$ defined in terms of matrix $w^{PK}$, see \eqref{tilder}  is related to
$r$ in (\ref{conmet}) via \cite{Candel}
\begin{equation}
	r^2 = \frac32 \,\widetilde{r}^{4/3}\,.
\label{rtilder}
\end{equation}

The equation of motion for $H_3$ following from the action \eqref{10Daction} reads
\begin{equation}
	d(e^{-\Phi}\ast H_3)= e^{-\Phi} d(\ast H_3) =0,
\label{H_3eqn}
\end{equation}
where $\ast $ denotes the Hodge star. 
In writing \eqref{H_3eqn} we neglected the back reaction on the dilaton to the leading order in small $H_3$ and assumed that the dilaton is constant (recall that this is the approximation of small $H_3$ that we assume in this Section; the back reaction will be included later in Sec.~\ref{sec:backreact}). Moreover, as we already mentioned  in the large $r$ limit,
 $r\gg |b|^{1/3}$ we will look for solutions of  \eqref{H_3eqn} using the metric \eqref{conmet} of the singular conifold.

For the 2-form $B_2$ we use the ansatz introduced in \cite{KlebStrass,KlebNekras,KlebTseytlin,tseytlin} for the type IIB ``compactifications'' on the conifold. 
 We start with a simple ansatz
\begin{equation}
	B_2= f_1(r)\, e_{\theta_1}\wedge e_{\varphi_1} + f_2(r)\, e_{\theta_2}\wedge e_{\varphi_2},
\label{B_2}
\end{equation}
where $f_1(r)$ and $f_2(r)$ are functions of the radial coordinate $r$, while the angle differentials are defined in \eqref{angles}. 
The 3-form field strength is then found as $H_3 = dB_2$,
\begin{equation}
	H_3= f_1' \, dr\wedge e_{\theta_1}\wedge e_{\varphi_1} + f_2' \, dr \wedge e_{\theta_2}\wedge e_{\varphi_2},
\label{H_3}
\end{equation}
where primes denote derivatives with respect to $r$.
Note that $dH_3$ trivially vanishes, so that the Bianchi identity is satisfied.

The 10D-dual of $H_3$ (calculated with respect to the metric \eqref{10metric}) is given by
\begin{equation}
	\ast H_3 =  \frac{1}{3}\, d \, {\rm Vol_4}  \wedge e_{\psi} \wedge \left( e_{\theta_2}\wedge e_{\varphi_2}\, r f_1'
	+ e_{\theta_1}\wedge e_{\varphi_1}\, r f_2'\right),
\label{astH_3}
\end{equation}
where $  d \, {\rm Vol_4} =dx^0 \wedge dx^1 \wedge dx^2 \wedge dx^3$ is the volume form for the 4D Minkowski spacetime.
Substituting $\ast H_3 $ into equation of motion \eqref{H_3eqn} we find the equations for the profile functions $f_1(r)$ and $f_2(r)$ from \eqref{B_2},
\begin{equation}
	(r\,f_1')'=0 \,, \quad
	(r\,f_2')'=0 \,, \quad
	f_1'+ f_2' =0
\end{equation}
First two of these equations come from the exterior derivative acting on $r f_1'$ and $r f_1'$, while the last one comes from acting with $d$ on $e_{\psi}$.
The solution to these equations has the form 
\begin{equation}
	f_1 = -f_2 = \mu_1 \log{r},
\label{f}
\end{equation}
where $\mu_1$ is a small real parameter.
Therefore, we can write the solution for $H_3$ as \cite{Y_NSflux}
\begin{equation}
	H_3=   \mu_1\, \alpha_3,
\label{H_3solution_1_alpha}
\end{equation}
where $\alpha_3$ is a real 3-form  defined as 
\begin{equation}
	\alpha_3 = \frac{dr}{r}\wedge \left(e_{\theta_1}\wedge e_{\varphi_1} -   e_{\theta_2}\wedge e_{\varphi_2}\right).
\label{alpha}
\end{equation}

As was noted in \cite{Y_NSflux}, there is one more independent solution for $H_3$.
Consider another 3-form on the conifold,
\begin{equation}
	\beta_3\equiv e_{\psi}\wedge \left(e_{\theta_1}\wedge e_{\varphi_1} -   e_{\theta_2}\wedge e_{\varphi_2}\right).
\label{beta}
\end{equation}
Both forms $\alpha_3$ and $\beta_3$ are closed \cite{KlebNekras,KlebTseytlin}.
Moreover, their 10D-duals are given by \cite{Y_NSflux}
\beq
\ast \alpha_3 =  -\frac{1}{3}\, d \, {\rm Vol_4}  \wedge \beta_3, \qquad \ast \beta_3 =  3\, d \, {\rm Vol_4}  \wedge \alpha_3,
\label{ast_alpha_beta}
\eeq
where the first relation was already shown in \eqref{astH_3}.
Therefore,  they satisfy also equations of motion
\begin{equation}
	d\ast\alpha_3 =0, \qquad d\ast\beta_3 =0.
\label{eqnofmotion}
\end{equation}
Hence, we can write the second solution for $H_3$,
\begin{equation}
	H_3 =\frac{\mu_2}{3} \,\beta_3,
\label{H_3solution_2_beta}
\end{equation}
where $\mu_2$ is another real parameter, while the factor $1/3$ is introduced for convenience. 
It follows that this $H_3$-form  satisfies both the Bianchi identity and equations of motion \eqref{H_3eqn}.

\vspace{10pt}

One may ask whether there are other solutions for the $H_3$ form.
To answer this question, note that since the $H_3$ form is closed, $d H_3 = 0$, it must belong to the third cohomology group $\mathcal{H}^3$ of the conifold.

For a 3-sphere we have $\mathcal{H}^3(S^3, \mathbb{Z}) \simeq \mathbb{Z}$ which is one-dimensional.
Our conifold is basically a product of $\mathbb{R}_{>0}$ and a fiber bundle of $\mathbb{S}^1$ over $\mathbb{S}^2 \times \mathbb{S}^2$.
Naively one might then expect two solutions for the $H_3$ form corresponding to two possible compact cycles $\mathbb{S}^3 \simeq \mathbb{S}^1 \times \mathbb{S}^2$ with different $\mathbb{S}^2$'s.
However, the equation of motion \eqref{H_3eqn} imposes an additional constraint, and only one combination of these two cohomology elements satisfies this equation, namely \eqref{beta}.
Consider now two other possible  non-compact 3-cycles $\mathbb{R}_{>0} \times \mathbb{S}^2$ with different $\mathbb{S}^2$'s.
Again, the equation of motion \eqref{H_3eqn} imposes an additional constraint, and only one combination survives. Thus, we do not expect other solutions for the $H_3$ form.

 Two 3-cycles associated with harmonic 3-forms $\alpha_3$  and $\beta_3$ represent a basis of two dual $A$ and $B$ cycles  on the conifold, which intersect each other only once \cite{Strom,GiddKachruPolch}. Due to non-compactness of the conifold  this basis of 3-forms is only log-normalizable ( see also  \cite{Y_NSflux}),
\beq
\int_{Y_6} \alpha_3 \wedge \beta_3 \sim -\int \frac{dr}{r} \sim - \log{\frac{R_{\rm IR}^3}{|b|}}. 
\label{ab}
\eeq
Here $R_{\rm IR}$ is the maximal value of the radial coordinate $r$ introduced to regularize the infrared logarithmic divergence,
while at small $r$ the integral is cut off by the minimal size of $S_3$ which is equal to $|b|$.  Note that this logarithm is similar to the one, which determines the metric
for the $b$-baryon in \eqref{Sb} \footnote{Note that $R_{\rm IR}^3\sim \widetilde{R}_{\rm IR}^2$, see \eqref{rtilder}.}. This logarithmic behavior will play an important role below.

\subsection{Deformation of the conifold} 
\label{sec:def}

Now let us reinstate $b$-dependence of the metric and compute the $H_3$ form in the fixed deformed conifold background.
For this purpose we can utilize the known expression for the exact metric on the deformed conifold \cite{Candel,Ohta,KlebStrass}.
This metric, parametrized by the deformation parameter $|b|$, is given by
\begin{equation}
\begin{aligned}
	ds_6^2 
		= \frac12\,|b|^{2/3} & K(\tau) \Bigg\{ \frac{1}{3K^3(\tau)}\left(d \tau^2 + e_{\psi}^2\right)  \\
		&+ \cosh^2{\frac{\tau}{2}}\,\left(g_3^2+ g_4^2\right) + \sinh^2{\frac{\tau}{2}}\,\left(g_1^2+ g_2^2\right) \Bigg\}
\end{aligned}
\label{defconmet}
\end{equation}
The new angle differentials are
\begin{equation}
\begin{aligned}
	g_1 &= -\frac1{\sqrt{2}}\,(e_{\phi_1} + e_3), \qquad &g_2 &= \frac1{\sqrt{2}}\,(e_{\theta_1} - e_4) \,, \\
	g_3 &= -\frac1{\sqrt{2}}\,(e_{\phi_1} - e_3), \qquad &g_4 &= \frac1{\sqrt{2}}\,(e_{\theta_1} + e_4) \,,
\end{aligned}
\label{g_angles}
\end{equation}
where
\begin{equation}
	e_3= \cos{\psi}\sin{\theta_2}\, d\varphi_2 - \sin{\psi}\,d\theta_2, \qquad 
	e_4= \sin{\psi}\sin{\theta_2}\, d\varphi_2 + \cos{\psi}\,d\theta_2.
\end{equation}
while $e_{\phi_{1,2}}$ and $e_{\theta_{1,2}}$ are the undeformed differentials from eq.~\eqref{angles}.
The angles here are defined at $0\le \theta_{1,2}<\pi$, $0\le \varphi_{1,2}<2\pi$, $0\le \psi<4\pi$.
Moreover, the $K(\tau)$ function in eq.~\eqref{defconmet} is defined as
\begin{equation}
	K(\tau)=\frac{(\sinh{2 \tau}-2 \tau)^{1/3}}{2^{1/3}\sinh{\tau}}
\end{equation}
The new radial coordinate $\tau$ is related to the $r$ of the singular cone via
\begin{equation}
	\widetilde{r}^2=|b|\,\cosh{\tau} = \left(\frac23\right)^{3/2}\,r^3.
\label{tau}
\end{equation}
Note that $\tau \in [0,\infty )$, and now there is a lower cutoff on $r$ given by $|b|^{1/3} \lesssim r$.
The square root of the metric determinant is computed to be
\begin{equation}
	\sqrt{g_6} = \frac{1}{96} |b|^2 \sin(\theta_1) \sin(\theta_2) \sinh[2](\tau)
\label{det_g6_nowarp}
\end{equation}

In the limit $\tau \gg 1$ (or equivalently $r \gg |b|^{1/3}$) the metric \eqref{defconmet} reduces to the metric of singular conifold \eqref{conmet}. 
In the opposite limit $\tau \ll 1$ the deformed conifold metric takes the form
\begin{equation}
	ds_6^2|_{\tau\to 0} 
		= \frac12\,|b|^{2/3}\left(\frac23\right)^{\frac13}\, \left\{ 
			\frac12\,d \tau^2 + \frac12\,e_{\psi}^2
			+ g_3^2+ g_4^2 +
			\frac{\tau^2}{4}\,\left(g_1^2+ g_2^2\right)
		\right\} \,.
\label{conmettau0}
\end{equation}
%

Our task is to compute the $H_3$ form in the fixed deformed conifold background.
The small and large $\tau$ limits were considered in \cite{Y_NSflux,NS}.
Here we will find  the exact solution and confirm that it behaves smoothly in the whole range of $\tau$.  Moreover, its small and  large-$\tau$ behavior coincide with one found in \cite{Y_NSflux,NS} .
Much in the same way as in Sec.~\ref{sec:H3_large_r}, we will find two independent solutions.

\subsubsection{First solution for the 3-form and parameter $\mu_1$}

To find the first solution generalizing eq.~\eqref{H_3solution_1_alpha} and eq.~\eqref{alpha}, we following \cite{Y_NSflux} start by writing the ansatz suggested in \cite{KlebStrass} for the type IIB flux compactification on the deformed conifold.
Writing $B_2 =  p(\tau)\,  g_1\wedge g_2 + k(\tau)\, g_3\wedge g_4 $ we get
\begin{equation}
	H_3 = p'(\tau)\, d\tau\wedge g_1\wedge g_2 + k'(\tau)\, d\tau\wedge g_3\wedge g_4 - \frac12\,[p(\tau) - k(\tau)]\, e_{\psi}\wedge 
	(g_1\wedge g_3 + g_2\wedge g_4),
\label{H_3tau1}
\end{equation}
where $p(\tau)$ and $k(\tau)$ are profile functions to be found below.
The primes denote derivatives with respect to $\tau$.
This 3-form is closed so  the Bianchi identity is satisfied. 
The 10D-dual of $ H_3$ reads
\begin{equation}
\begin{aligned}
	\ast H_3 =  \,d \, {\rm Vol_4} \, \wedge \, \Bigg( & p^{\prime} \frac{1}{\tanh[2](\frac{\tau}{2})} \, e_{\psi} \wedge g_3 \wedge g_4 
			+ k^{\prime} \tanh[2]( \frac{\tau}{2} ) \,  e_{\psi} \wedge g_1 \wedge g_2 \\
		&- \frac{1}{2}(p-k)  \, d \tau \wedge (g_1 \wedge g_3 + g_2 \wedge g_4) \Bigg),
\end{aligned}
\end{equation}
Substituting $\ast H_3$ into the equation of motion \eqref{H_3eqn}  and using the  identity  \cite{KlebStrass}
\beq
d(g_1\wedge g_3 + g_2\wedge g_4) =  e_{\psi}\wedge (g_1\wedge g_2 - g_3\wedge g_4)
\label{id2}
\eeq
 we find two equations,
\begin{equation}
\begin{aligned}
	\partial_{\tau} \left( \frac{p^{\prime}}{\tanh[2](\frac{\tau}{2})} \right) - \frac{1}{2}(p-k) &= 0 \,, \\
	\partial_{\tau} \left( k^{\prime} \tanh^2 \left(\frac{\tau}{2} \right) \right) + \frac{1}{2} (p-k) &= 0 \,.
\end{aligned}	
\label{eq_for_p}
\end{equation}
Asymptotics of the profile functions $p(\tau)$ and $k(\tau)$ was found in \cite{Y_NSflux} in the limit of small  $\tau$.
Here we solve these  equations  exactly.

From the sum of  two equations \eqref{eq_for_p}  we easily find
\begin{equation}
	p^{\prime} = \tilde{\mu}_1 \tanh^2 \left(\frac{\tau}{2} \right)  - k^{\prime} \tanh^4 \left(\frac{\tau}{2}\right), 
\label{eq_for_p'}
\end{equation}
where $\tilde{\mu}_1$ is an integration constant. 
From the second equation in \eqref{eq_for_p} we can express $p$,
\begin{equation}
	p = k - 2  \partial_{\tau} \Big( k^{\prime} \tanh[2](\frac{\tau}{2}) \Big).
\label{p}
\end{equation}
Let us define
\begin{equation}
	u = k^{\prime} \tanh[2](\frac{\tau}{2}).
\label{u_eq}
\end{equation}
Substituting $p$ from eq.~\eqref{p} into \eqref{eq_for_p'} and using \eqref{u_eq} we obtain the following second order equation for $u(\tau)$:
\begin{equation}
	- u''+ \frac{u}{2} \left( \frac{1}{\tanh[2](\frac{\tau}{2}) } + \tanh[2](\frac{\tau}{2}) \right) = \frac{\tilde{\mu}_1}{2} \tanh[2](\frac{\tau}{2}). 
\label{eq_for_u}
\end{equation}

One can explicitly check that the  solutions\footnote{One can arrive at the first of these solutions by noting that eq.~\eqref{eq_for_u} at $\tilde{\mu}_1=0$ resembles the Schr\"odinger equation with the P\"oschl-Teller potential with the coordinate shifted by $i\pi/2$. After that, one can do a reduction to a first order equation and find the second solution.} of the homogeneous version of that equation (i.e. for $\widetilde{\mu}_2 = 0$) are
\begin{equation}
\begin{aligned}
	u_0^{(1)}(\tau) &= \frac{c_1}{ \sinh(\tau) } \\
	u_0^{(2)}(\tau) &= \frac{c_2}{ \sinh(\tau) } \left( -\frac{\tau}{2} + \frac{1}{4} \sinh(2\tau) \right),
\end{aligned}
\label{u_homog}
\end{equation}
where $c_1$ and $c_2$ are arbitrary constants.
By using the explicit form of these homogeneous solutions one can easily find the solution of the original inhomogeneous equation \eqref{eq_for_l} that behaves smoothly for all $\tau$,
\begin{equation}
    u(\tau) = \frac{\widetilde{\mu}_1}{2} \left( 1 -  \frac{\tau}{\sinh(\tau)} \right) \,.
\label{u}
\end{equation}
The general solution of equation \eqref{eq_for_u} is given by the sum of the solution \eqref{u} and two solutions \eqref{u_homog} of the homogeneous equation.  Observe now that the first solution in  \eqref{u_homog} is singular at $\tau\to 0$, while the second one is singular at $\tau\to\infty$. To avoid singularities we take both solutions of the homogeneous equation with zero coefficients, $c_1=c_2=0$. This leads to the solution \eqref{u}. 

Using the relations \eqref{eq_for_p'} and \eqref{u_eq} we can rewrite the expression for the $H_3$ form, eq.~\eqref{H_3tau1}, as
\begin{equation}
	H_3 = \tanh[2](\frac{\tau}{2})[\tilde{\mu}_1  - u(\tau) ]\, d\tau \wedge g_1\wedge g_2 
		+ \frac{u(\tau) }{\tanh[2](\frac{\tau}{2})} \, d\tau\wedge g_3\wedge g_4 
		+ u^{\prime}(\tau)  e_{\psi} \wedge (g_1\wedge g_3 + g_2\wedge g_4)
\label{H_3_first}
\end{equation}
where the profile function $u(\tau)$ is given by eq.~\eqref{u}.
In the limit of large $\tau$ $u\approx  \tilde{\mu}_1/2$ so the large-$\tau$ behavior of \eqref{H_3_first} is simply
\begin{equation}
	H_3 \approx \frac{\tilde{\mu}_1}{2} \, d\tau\wedge \left(g_1\wedge g_2 \,+\,  g_3\wedge g_4\right)
\label{H_3_first_large_tau}
\end{equation}
Noting the relation $d\tau = 3 dr / r$ following from \eqref{tau} at large $\tau$ and identity  \cite{KlebStrass}
\beq
e_{\theta_1}\wedge e_{\varphi_1} - e_{\theta_2}\wedge e_{\varphi_2}= g_1\wedge g_2 + g_3\wedge g_4
\label{id0}
\eeq
 we see that \eqref{H_3_first_large_tau} exactly reproduces the large-$r$ asymptotic \eqref{H_3solution_1_alpha} if we identify
\begin{equation}
	\tilde{\mu}_1 = \frac{2}{3}\, \mu_1
\label{mu1_tilda_identification}
\end{equation}
In the opposite limit of small $\tau\;$ $u\approx  \tilde{\mu}_1\,\tau^2/12$ we get
\begin{equation}
	H_3 \approx \tilde{\mu}_1 \frac{\tau^2}{4} \, d\tau\wedge g_1\wedge g_2 
		+ \frac{\tilde{\mu}_1}{3} \, d\tau\wedge g_3\wedge g_4 
		+ \tilde{\mu}_1 \frac{ \tau }{6}\, e_{\psi}\wedge (g_1\wedge g_3 + g_2\wedge g_4),
\label{H_3_first_small_tau}
\end{equation}
where the second term dominates in the limit $\tau\to 0$. The solution for $H_3$ in this limit was calculated in \cite{Y_NSflux}.
The exact solution in Eqs.~\eqref{H_3_first}, \eqref{u} demonstrates that the small-$\tau$ asymptotic, \eqref{H_3_first_small_tau}, is smoothly connected to the large-$\tau$ asymptotic, \eqref{H_3_first_large_tau}. This close the gap in \cite{Y_NSflux} where this question was not clarified. Say, in principle, it could happen that the solution for $H_3$ with the  large-$r$ asymptotic \eqref{H_3solution_1_alpha}
develops a singularity at small $\tau$. Fortunately this does not happen.

To conclude this section, we note that at $\tau=0$ the  solution \eqref{H_3_first_small_tau}  tends to a constant
\beq
H_3(\tau=0) = \frac{ \tilde{\mu}_1}{3}\,d\tau\wedge g_3\wedge g_4,
\label{bc1}
\eeq
which we impose as a  boundary condition at $S_3$, which does not shrinks to zero at $\tau=0$.  This boundary condition ensures a non-zero solution for $H_3$.

\subsubsection{Second solution for the 3-form and parameter $\mu_2$}

To find the second solution for the $H_3$ form generalizing eq.~\eqref{H_3solution_2_beta} and eq.~\eqref{beta}, we use an ansatz \cite{NS} 
\begin{equation}
	H_3= l(\tau)\, e_{\psi}\wedge g_1\wedge g_2 + n(\tau)\, e_{\psi}\wedge g_3\wedge g_4 + q(\tau)\, d\tau\wedge (g_1\wedge g_3 + g_2\wedge g_4), 
\label{H_3tau2}
\end{equation}
where $l(\tau)$, $n(\tau)$ and $q(\tau)$ are profile functions to be determined.
The Bianchi identity $d H_3 =0$ implies
\begin{equation}
	l^{\prime}-q=0 \,, \quad
	n^{\prime}+q=0 \,,
\end{equation}
where we used the identity \eqref{id2}.
These equations are  solved by
\begin{equation}
	n = \widetilde{\mu}_2-l \,, \quad 
	q = l^{\prime} \,,
\label{n_q_via_l}
\end{equation}
where $\widetilde{\mu}_2$ is a new integration constant. 
The 10D-dual of $H_3$ from \eqref{H_3tau2} is found to be
\begin{equation}
\begin{aligned}
	\ast H_3  = - \,d \, {\rm Vol_4} \, \wedge \, \Bigg( & \frac{l}{\tanh[2](\frac{\tau}{2})} \, d \tau \wedge g_3 \wedge g_4
			+\left(\tilde{\mu}_2-l\right) \tanh[2](\frac{\tau}{2})  d \tau \wedge g_1 \wedge g_2 \\
		&+ l^{\prime} \, e_{\psi} \wedge\left(g_2 \wedge g_4+g_1 \wedge g_3\right) \Bigg) \,.
\end{aligned}
\end{equation}
Substituting this expression into the equation of motion \eqref{H_3eqn} we obtain a second order equation for $l(\tau)$:
\begin{equation}
    - l'' + \frac{l}{2}  \left( \frac{1}{\tanh[2](\frac{\tau}{2}) } + \tanh[2](\frac{\tau}{2}) \right) 
    	= \frac{\widetilde{\mu}_2}{2} \tanh[2](\frac{\tau}{2}) \,,
\label{eq_for_l}
\end{equation}
where we used the identity \cite{KlebStrass}
\beq
d(g_1\wedge g_2 - g_3\wedge g_4) = - e_{\psi}\wedge (g_1\wedge g_3 + g_2\wedge g_4).
\label{id1}
\eeq

Incidentally the equations for profile functions in the two cases, eq.~\eqref{eq_for_u} and eq.~\eqref{eq_for_l}, come out to be identical up to a replacement
\begin{equation}
	u(\tau) \leftrightarrow l(\tau) \,,  \quad 
	\tilde{\mu}_1 \leftrightarrow \tilde{\mu}_2 \,.
\label{corr}
\end{equation}
This allows us to immediately write down the solution for $l(\tau)$:
\begin{equation}
    l(\tau) = \frac{\widetilde{\mu}_2}{2} \left( 1 -  \frac{\tau}{\sinh(\tau)} \right) \,.
\label{l}
\end{equation}

Now that we have the profile function $l(\tau)$ at hand we can study the $H_3$ form.
Substituting \eqref{n_q_via_l} into the initial ansatz \eqref{H_3tau2} yields
\begin{equation}
	H_3= l(\tau)  \, e_{\psi}\wedge g_1\wedge g_2 + (\tilde{\mu}_2-l(\tau) )\, e_{\psi}\wedge g_3\wedge g_4 +l^{\prime}(\tau)  \, d\tau\wedge (g_1\wedge g_3 + g_2\wedge g_4),
\label{H_3_second}
\end{equation}
The function $l(\tau)$ is explicitly given by eq.~\eqref{l}.
At large $\tau$, eq.~\eqref{l} reduces to $l(\tau) \approx \widetilde{\mu}_2/2$.
With the help of the  identity \eqref{id0}
one can see that for large $\tau$ the $H_3$ form \eqref{H_3_second} simplifies to
\begin{equation}
	H_3 \approx \frac{\widetilde{\mu}_2}{2} e_\psi \wedge \Big( e_{\theta_1}\wedge e_{\varphi_1} - e_{\theta_2}\wedge e_{\varphi_2} \Big)
\end{equation}
This precisely reproduces the large-$r$ solution \eqref{H_3solution_2_beta} upon identification
\begin{equation}
	\widetilde{\mu}_2 = \frac{2}{3} \,\mu_2
\label{mu2_tilda_identification}
\end{equation}
At small $\tau$ we have $l(\tau) \approx \widetilde{\mu}_2 \, \tau^2 / 12$.
The $H_3$ form \eqref{H_3_second} becomes in this approximation
\begin{equation}
	H_3 \approx \frac{\widetilde{\mu}_2 \tau^2}{12}\,e_{\psi}\wedge g_1\wedge g_2 + \widetilde{\mu}_2 \,e_{\psi}\wedge g_3\wedge g_4 +\frac{\widetilde{\mu}_2 \tau}{6}\, 
	d\tau\wedge ( g_1\wedge g_3 +g_2\wedge g_4).
\label{2nd_sol_tau}
\end{equation}
Again, the small- and large-$\tau$ asymptotics smoothly interpolate into each other.

 Much in the same way as for the first solution the  solution \eqref{2nd_sol_tau}  tends to a constant  at $\tau=0$,
\beq
H_3(\tau=0) =  \tilde{\mu}_2\, e_{\psi}\wedge g_3\wedge g_4,
\label{bc2}
\eeq
which we impose as  a boundary condition at $S_3$  at $\tau=0$.

\subsection{Baryon potential $V(b)$}
\label{sec:nobackreaction_Vb}

Turning on the $H_3$ form flux induces a potential for the conifold complex structure  parameter $b$ (recall that $b$ becomes a scalar field in the 4D SQCD), since the action computed on the $H_3$ background flux is going to depend on $b$.
Let us review the derivation of the potential generated by the $H_3$ form \cite{Y_NSflux,NS}.

To obtain the potential we need to substitute the solution for the gravity background into the 10D action \eqref{10Daction}.
In fact, by using the equations of motion we can express the action integrand in terms of the $H_3$ form.
Einstein's equations following from \eqref{10Daction} read
\begin{equation}
	R_{MN}= \frac12 \,\pt_M\Phi\,\pt_N\Phi + \frac{e^{-\Phi}}{4}\, H_{MAB}H_N^{AB} - \frac{e^{-\Phi}}{48}\,G_{MN}\, (H_3)^2.
\label{Einstein}
\end{equation}
where $(H_3)^2 = H_{MNL} H^{MNL}$.
Taking the trace we find
\begin{equation}
	R - \frac12 G^{MN}\,\pt_M\Phi\,\pt_N\Phi -\frac{e^{-\Phi}}{24}\, (H_3)^2 =0.
\label{R_10}
\end{equation}
Substituting eq.~\eqref{R_10} back into the 10D action \eqref{10Daction} we arrive at
\begin{equation}
	S_{10D}= \frac1{2\kappa^2}\,\int d^{10} x\sqrt{-G}\left\{-\frac{e^{-\Phi}}{24}\, (H_3)^2 \right\}.
\end{equation}
The effective potential in the 4D Minkowski spacetime is then read off as
\begin{equation}
	V(b)=\frac{T^4}{(2\pi)^3g_s^2}\,\int d^6 x \sqrt{g_6}\;\frac{e^{-\Phi}}{24}\, (H_3)^2 ,
\label{pot_gen}
\end{equation}
where the $d^6 x$ integration is carried over the 6D Calabi-Yau manifold $Y_6$, and $g_6$ is the determinant of the 6D metric. 
In writing eq.~\eqref{pot_gen} we also used the expression for the gravitational coupling $\kappa$ in terms of the confining string tension $T$, see eq.~\eqref{kappa_T_relation}.

%
%

In \cite{Y_NSflux,NS}, an approximate solution for the $H_3$ form was found in the limits of small and large $\tau$.
From that one can derive an asymptotic form of the effective potential $V(b)$.

\begin{figure}[h]
    \centering
    \begin{subfigure}[t]{0.49\textwidth}
        \centering
        \includegraphics[width=\textwidth]{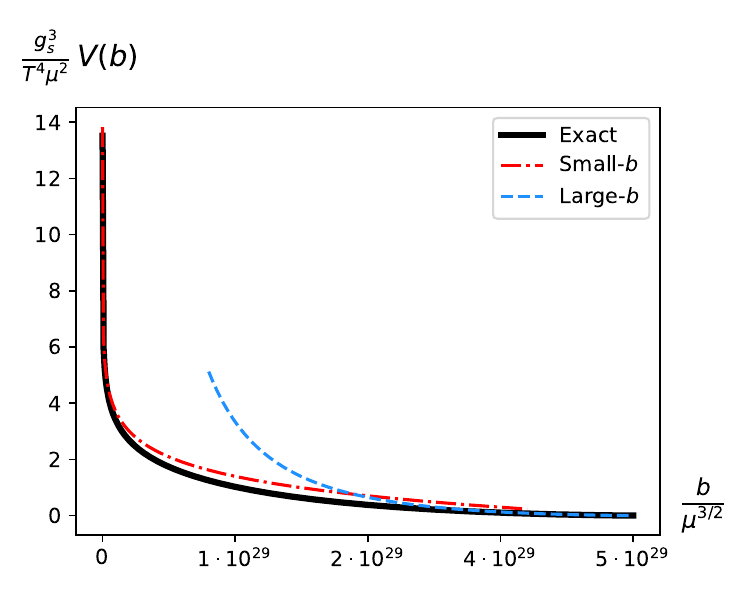}
        \caption{Linear $b$-scale}
        \label{fig:Veff_noback reaction_normalscale}
    \end{subfigure}%
    ~ 
    \begin{subfigure}[t]{0.49\textwidth}
        \centering
        \includegraphics[width=\textwidth]{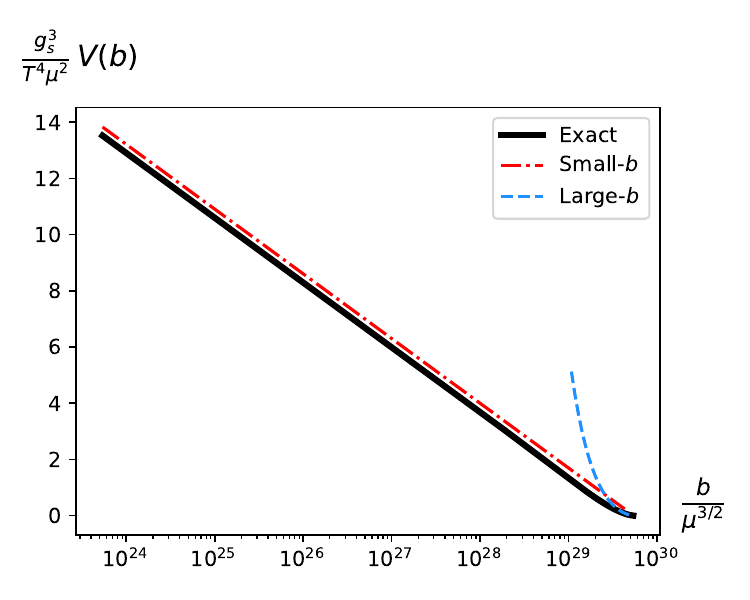}
        \caption{Log $b$-scale}
        \label{fig:Veff_noback reaction_logscale}
    \end{subfigure}%
	\caption{
		The baryon potential $V(b)$ generated by the $H_3$ form flux.
		Both plots represent the same data, with regular scaling of the $b$-axis on (\subref{fig:Veff_noback reaction_normalscale}) and logarithmic scaling on (\subref{fig:Veff_noback reaction_logscale}).
		The solid black lines represent the exact formula \eqref{Veff_noback reaction_full}, the red dash-dotted lines are the small-$b$ asymptotic, see eq.~\eqref{pot_log}, while the blue dashed lines correspond to the large-$b$ asymptotic, see eq.~\eqref{pot_b_to_b_IR}.
		One can see that the asymptotic formulas indeed give the approximations for the exact expression.
		The potential drives $b$ to large values and leads to a runaway vacuum.
		For this plot $\mu^2 \equiv (\tilde{\mu}_1^2 + \tilde{\mu}_2^2) = 1$, $R_\text{IR}=10^{10}$		
	}
\label{fig:Veff_noback reaction}
\end{figure}

Now we have the exact solution for the $H_3$ form valid for all $\tau$, so we are in a position to compute the exact potential as a function of  $b$ (but we still neglect the back reaction on the metric, i.e. we consider a fixed background metric on the deformed conifold).
To this end we need to substitute the solution for the $H_3$ form into eq.~\eqref{pot_gen}.
For the first solution \eqref{H_3_first} we have (note that $\tanh[4](\tau/2)$ is partially canceled by the metric components $g^{g_1 g_1}g^{g_2 g_2}$ and $g^{g_3g_3}g^{g_4g_4}$)
\begin{equation}
	\sqrt{g_6} (H_3)^2 = 3! \left( \frac{u(\tau)^2}{ \tanh[2](\frac{\tau}{2})} + (\tilde{\mu}_1-u(\tau))^2  \tanh[2](\frac{\tau}{2}) + 2 u'(\tau)^2 \right)
\label{H3_square_1}
\end{equation}
The second solution \eqref{H_3_second} gives (note that  $ \tanh[2](\tau/2)$ appears due to the factors $g^{g_1 g_1}g^{g_2 g_2}$ and $g^{g_3g_3}g^{g_4g_4}$)
\begin{equation}
	\sqrt{g_6} (H_3)^2 = 3! \left( \frac{l(\tau)^2}{ \tanh[2](\frac{\tau}{2})} + (\tilde{\mu}_2-l(\tau))^2  \tanh[2](\frac{\tau}{2}) + 2 l'(\tau)^2 \right)
\label{H3_square_2}
\end{equation}
Substituting $(H_3)^2$ from eq.~\eqref{H3_square_1} and eq.~\eqref{H3_square_2} into eq.~\eqref{pot_gen} yields the exact potential.
The integral over the radial coordinate $\tau$ on the deformed conifold is IR divergent, so we impose a cutoff, $0 \leqslant \tau \leqslant \tau_\text{IR}$.
The integration can be performed analytically with the result  
\begin{subequations}
\begin{equation}
	V(b) = \frac{T^4}{g_s^3}\, (\tilde{\mu}_1^2 +\tilde{\mu}_2^2) \, \left(
		\tau_\text{IR}
		- \frac{ \cosh(\tau_\text{IR}) }{ \sinh(\tau_\text{IR}) }
		+ \frac{ 2 \tau_\text{IR} }{ \sinh[2](\tau_\text{IR}) }
		- \frac{ \tau_\text{IR}^2 \cosh(\tau_\text{IR}) }{ \sinh[3](\tau_\text{IR}) }
		\right)
\label{Veff_nobackreaction}
\end{equation}
\begin{equation}
	|b|\cosh(\tau_\text{IR}) = \left(\frac23\right)^{3/2} R_\text{IR}^3
\label{tauIR_matching}
\end{equation}
\label{Veff_noback reaction_b-tau}
\end{subequations}
for the  potential $V (b)$  induced by two solutions for $H_3$, \eqref{H_3_first} and \eqref{H_3_second} associated with two parameters $ \tilde{\mu}_1$ and $ \tilde{\mu}_2)$ \footnote{Note that there is no overlap between two $H_3$ forms.}.
The relation between the IR cutoffs in terms of $\tau$ and $r$, eq.~\eqref{tauIR_matching}, is coming from the general relation \eqref{tau}.
$V(b)$ in eq.~\eqref{Veff_nobackreaction} depends on $b$ through $\tau_\text{IR}$ in \eqref{tauIR_matching}.
Note, that  dimensions of parameters $\tilde{\mu}_i$ are $({\rm mass})^{-2}$, $i=1,2$.

Substituting $b$ from eq.~\eqref{tauIR_matching} into eq.~\eqref{Veff_nobackreaction} we finally arrive at
\begin{equation}
	V(b) = \frac{T^4}{g_s^3}\, (\tilde{\mu}_1^2 + \tilde{\mu}_2^2)\,
	\frac{ (|b|^2+b_{\text{IR}}^2 )\, {\rm arccosh}( \frac{b_{\text{IR}}}{|b|} ) }{ b_{\text{IR}}^2 - |b|^2}
	-\frac{ |b|^2 b_{\text{IR}} \left({\rm arccosh}( \frac{b_{\text{IR}}}{|b|} )^2 - 1 \right) + b_{\text{IR}}^3 }{ (b_{\text{IR}}^2-|b|^2)^{3/2}}
\label{Veff_noback reaction_full}	
\end{equation}
where $b_{\text{IR}} = (2/3)^{3/2} R_\text{IR}^3$ is now a fixed parameter of IR regularization in terms of conifold radial coordinate $r$. 

In the small  $|b|$ limit the result  \eqref{Veff_noback reaction_full}  gives 
\beq
V(b) = \frac{T^4 }{g_s^3}  (\tilde{\mu}_1^2 + \tilde{\mu}_2^2) \log{ \dfrac{b_{\text{IR}}}{  |b|}}   \quad  
			\text{for small } |b| \ll R_\text{IR}^3
\label{pot_log}
\eeq

This result was obtained in \cite{Y_NSflux}. Note that writing
\beq
\int d^{10} x\sqrt{-G}\, (H_3)^2 \sim \int H_3\wedge \ast H_3
\eeq
and substituting solutions \eqref{H_3solution_1_alpha}, \eqref{H_3solution_2_beta} for $H_3$  we get the same logarithmic behavior of the potential using \eqref{ab}. The presence of the same logarithmic factor in the metric \eqref{Sb} for the b-baryon is consistent with \ntwo supersymmetry in 4D theory, see  \cite{Louis2} for  CY compactifications with NS 3-form flux
 in the type IIA string theory.

We see that the Higgs branch for the modulus $b$ is lifted  and  the  potential in  \eqref{Veff_noback reaction_full} leads to  the runaway vacuum, see Fig.~\ref{fig:Veff_noback reaction}, in particular, it does not generate a mass term for $b$ \cite{Y_NSflux}.

When $b$ approaches $b_{\text{IR}} = (2/3)^{3/2} R_\text{IR}^3$  the potential \eqref{Veff_nobackreaction} 
 vanishes as
\begin{equation}
	V (b) \approx \dfrac{2^{5/2}}{9} \dfrac{T^4}{g_s^3}\, (\tilde{\mu}_1^2 + \tilde{\mu}_2^2)
 \,\left( \frac{   b_{\text{IR}} - |b| }{| b| } \right)^{3/2}
\label{pot_b_to_b_IR}
\end{equation}
This behavior was obtained in \cite{Y_NSflux}.  We see that to minimize  the potential above $|b|$ becomes large and approaches the infrared cutoff,
\beq
 <|b| > =  b_{\text{IR}} \to \infty.
\label{VEVb}
\eeq

In fact, in the limit $|b|\to b_{\text{IR}}$  the potential  \eqref{pot_b_to_b_IR} is proportional to $\tau_\text{IR}^3$ which is  the volume of the  three dimensional ball bounded by the  sphere $S_2$ of the conifold with the maximum radius $\tau_\text{IR}$. It shrinks to zero as $b$ tends to its VEV in \eqref{VEVb}.
To avoid singularities we can regularize the size of $S_2$  introducing small non-zero $\beta$, which  makes the conifold ''slightly resolved'', see \eqref{D-term}. We  take the limit $\beta\to 0$ at the last step. Then the value of the potential and all its derivatives vanish in the  vacuum \eqref{VEVb} at $<|b|>=b_{\text{IR}}  $, for example 
\beq
V (|b|=b_{\text{IR}} )
 \sim  (\tilde{\mu}_1^2 + \tilde{\mu}_2^2) \, \frac{T^4}{g_s^3}\,  \frac{\beta^3}{R_{\rm IR}^{9/2}} \to 0.
\label{VatVEV}
\eeq

Vanishing of the  potential $V (b)$ together with all its derivatives at  the runaway vacuum confirms that \ntwo supersymmetry is not broken in 4D SQCD.

In the next Section we are going to include the effect of back reaction of the $H_3$ flux on the metric and the dilaton.

\section{ Including back reaction }
\label{sec:backreact}

In the discussion above we constrained ourselves to the limit $|b|^{2/3} \gg \mu_1, \mu_2$ when the $H_3$ flux is small.
Relaxing this constraint requires one to take into account  back reaction of the flux on the spacetime metric and the dilaton.
This can be done by introducing warp factors into the conifold metric and solving the resulting gravity equations of motion.

Now we concentrate on the limit 
\beq
|b|^{2/3} \ll \mu_1, \mu_2.
\label{large_mu}
\eeq
Anticipating that the change of the regime could appear at $r^2\sim \mu_1, \mu_2$ we will work with the conifold region $r \gg |b|^{1/3}$.  In this region  we can use the metric of the singular conifold; as is discussed below, this is sufficient for our purposes.
We stress that this time we do not require $r^2$ to be large compared to the flux parameters $\mu_1,\mu_2$.
We will see that behavior of the conifold metric changes when $r^2$ goes below $\mu_1,\mu_2$.

We start this Section by discussing the warp compactification setup and then discuss the solution for the warp factors and implications on the baryon potential $V(b)$.

\subsection{Warp factors}

Following \cite{NS}, we start by introducing warp factors into the 10D metric \eqref{10metric}, \eqref{conmet}.
Our ansatz is
\begin{subequations}
\begin{equation}
	ds^2_{10} = h^{-1/2}_4 (r) \,\eta_{\mu\nu} dx^{\mu}dx^{\nu}+ \,g_{mn} dx^m dx^n,
\label{10met}
\end{equation}
\begin{equation}
	g_{mn} dx^m dx^n = h^{1/2}_6 (r) \left\{ 
			a(r)\,dr^2 
			+ \frac{r^2}{6}(e_{\theta_1}^2+ e_{\varphi_1}^2 +e_{\theta_2}^2+ e_{\varphi_2}^2) 
			+\frac{r^2}{9} \omega(r)\,e_{\psi}^2
		\right\},
\label{warpedconi}
\end{equation}
\label{warped_singular_coni_full_metric}
\end{subequations}
Here, $\mu,\nu =0,...,3$ are indices of the 4D space and $\eta_{\mu\nu}$ is the  flat Minkowski metric with signature $(-1,1,1,1)$, while $m,n= 5,...10$ are indices  of the 6D internal space. 
We assume that the warp factors $h_4$, $h_6$, $a$ and $\omega$ depend only on the radial conifold coordinate $r$.
Generalization of the volume form \eqref{Vol} in this case reads
\begin{equation}
	({\rm Vol})_\text{warped 10D} 
		=  \int d^4 x
		\frac{1}{108} \int \frac{1}{h_4} \, a^{1/2} \omega^{1/2} h_6^{3/2} \, r^5 \,dr \, d\psi\,  d\theta_1\,\sin{\theta_1} d\varphi_1 \,d\theta_2 \, \sin{\theta_2}d\varphi_2 \,.
\label{Vol_warped}
\end{equation}
The 4D measure $d^4 x$ corresponds to the Minkowski space.

\subsection{Dilaton and the NS 3-form}

The bosonic part of the action of the type IIA supergravity is given in Eq. \eqref{10Daction}. As mentioned previously, the equation of motion for $H_3$ following from action reads 
\beq \label{eq_for_H3}
d(e^{-\Phi}\ast H_3) =0.
\eeq
Einstein's equations of motion following from action \eqref{10Daction} have the form
\beq
R_{MN}= \frac12 \,\pt_M\Phi\,\pt_N\Phi + \frac{e^{-\Phi}}{4}\, H_{MAB}H_N^{AB} - \frac{e^{-\Phi}}{48}\,G_{MN}\, H_3^2,
\eeq
while the equation for the dilaton reads
\beq
G^{MN}D_M\Phi D_N \Phi + \frac{e^{-\Phi}}{12}\, H_3^2 =0.
\label{dilatoneq}
\eeq
In \cite{NS}, it was shown that the Einstein equations for Minkowski indices and the dilaton equation of motion are equivalent upon substitution
\beq 
\Phi = \Phi_0 + \ln{h_4},
\label{dilaton_solution}
\eeq
where $\Phi_0$ is a constant value of the dilaton present at $H_3=0$.

Our next task is to solve the equation of motion \eqref{eq_for_H3} for $H_3$ in the warped metric.
As previously, we find two possibilities corresponding to two different solutions.

For the first solution, we start with the following expression (cf. eq.~\eqref{H_3}) $H_3 = dB_2$
\begin{equation}
    H_3 = f_1' dr \wedge e_{\theta_1} \wedge e_{\varphi_1} + f_2' dr \wedge e_{\theta_2} \wedge e_{\varphi_2} 
\end{equation}
Here, $f_1(r)$ and $f_2(r)$ are unknown profile functions, and primes denote derivatives with respect to $r$.
The $H_3$ form thus defined is closed and satisfies the Bianchi identity.
Computing it's Hodge dual we obtain
\begin{equation}
	e^{-\Phi} \ast H_3 
		= \frac{1}{3} \frac{\omega^{1/2} e^{-\Phi}}{a^{1/2} h_4} \, d {\rm Vol_4} \wedge e_{\psi} \wedge 
		\left( e_{\theta_2}\wedge e_{\varphi_2}\, r f_1' + e_{\theta_1}\wedge e_{\varphi_1}\, r f_2'\right)
\end{equation}
As above, the 4D Minkowski volume form is simply $d \, {\rm Vol_4} = d x_0 \wedge dx_1 \wedge dx_2 \wedge dx_3$.
Substituting this into equation of motion \eqref{eq_for_H3} yields equations for the profile functions $f_1(r)$ and $f_2(r)$,
\begin{equation}
\begin{aligned}
	\partial_r \frac{e^{-\Phi} \omega^{1 / 2} r f_1^{\prime}}{a^{1 / 2} h_4} &= 0 \\
	\partial_r \frac{e^{-\Phi} \omega^{1 / 2} r f_2^{\prime}}{a^{1 / 2} h_4} &= 0 \\
	\frac{e^{-\Phi} \omega^{1 / 2}}{a^{1/2} h_4} \left(f_1^{\prime}+f_2^{\prime}\right) &= 0
\end{aligned}
\end{equation}
The third equation gives $f_1'=-f_2'$, while the first equation above together with the solution \eqref{dilaton_solution} 
for the dilaton taken into account yields
\begin{equation}
    f_1' = \frac{\mu_1}{r} \frac{h_4^2 a^{1/2}}{\omega^{1/2}},
\end{equation}
where $\mu_1$ is an integration constant.
Thus we obtain for the $H_3$ form
\begin{equation}
	H_3 = \mu_1  \frac{h_4^{2} a^{1/2} }{\omega^{1/2}} \frac{d r}{r} \wedge\left(e_{\theta_1} \wedge e_{\varphi_1}-e_{\theta_2} \wedge e_{\varphi_2}\right)
\label{H3_first_solution_warped}
\end{equation}
In the no-back-reaction limit when all the warp factor go to unity, eq.~\eqref{H3_first_solution_warped} reproduces eq.~\eqref{H_3solution_1_alpha}.

To find the second solution, we can start from the same ansatz as in the case with no warp factors, eq.~\eqref{H_3solution_2_beta}
\begin{equation} 
    H_3 = \frac{\mu_2}{3} e_\psi \wedge (e_{\theta_1} \wedge e_{\varphi_1} - e_{\theta_2} \wedge e_{\varphi_2})
\label{H3_second_solution_warped}
\end{equation}
where $\mu_2$ is a parameter. 
This form is closed, so the Bianchi identity is satisfied. 
Computing the Hodge dual we obtain
\begin{equation}
	e^{-\Phi} \ast H_3 = \mu_2 \frac{a^{1/2} e^{-\Phi}}{\omega^{1/2} h_4} \, d {\rm Vol_4} \wedge \frac{dr}{r}   \left(e_{\theta_1}\wedge e_{\varphi_1} -   e_{\theta_2}\wedge e_{\varphi_2}\right)
\label{H3_second_solution_warped_dual}
\end{equation}
Since the warp factors and the dilaton depend only on $r$, this form is also closed, and the equation \eqref{eq_for_H3} is trivially satisfied.

Below we study the Einstein equations for the warp factors induced by $H_3$ flux. Then we consider the  resulting baryon potential $V(b)$.

\subsection{Baryon potential from $\mu_2$ flux}

Let us start with discussing the second solution for the $H_3$ form parametrized by $\mu_2$, eq.~\eqref{H3_second_solution_warped}, as this case turns out to be more tractable.

\subsubsection{Einstein equations}

In order to study the effect that the $H_3$ form has on the metric, we need to derive Einstein equations for the metric.
By using the the metric in the form \eqref{warped_singular_coni_full_metric}, we can compute the Christoffel symbols and the Ricci tensor in terms of the warp factors.
This technical part was done in Wolfram Mathematica \cite{git_conifold_einstein}.
The results can be found in Appendix~\ref{sec:warped_sing_ricci}.

Because of the two different solutions for the $H_3$ form (see eq.~\eqref{H3_first_solution_warped} and eq.~\eqref{H3_second_solution_warped}), there are two possibilities for the $H_3$ flux.
The corresponding Einstein equations are derived in Appendix~\ref{sec:warped_sing_ricci}.
In each of the two cases they will give different ODEs for the warp factors.
For the second solution for $H_3$, eq.~\eqref{H3_second_solution_warped}, by taking linear combinations of Einstein equations in eq.~\eqref{ricci_einstein_eq_mu2} we can bring the resulting system of equations to the form
\begin{subequations}
\begin{equation}
	\frac{h_4'' }{h_4 } 
	-\frac{20}{r^2}+\frac{24 a }{r^2}-\frac{4 a  \omega  }{r^2}-\frac{4 \omega ' }{r \omega  }+\frac{15
	h_4' }{r h_4 }-\frac{a'  h_4' }{2 a  h_4 }+\frac{3 \omega '  h_4' }{2 \omega   h_4 }-\frac{9 h_4' {}^2}{4 h_4 {}^2}-\frac{10
	h_6' }{r h_6 }-\frac{\omega '  h_6' }{\omega   h_6 }+\frac{7 h_4'  h_6' }{2 h_4  h_6 }-\frac{5 h_6' {}^2}{4 h_6 {}^2}
	=0
\label{einstein_eq_mu2_h4}
\end{equation}
\begin{equation}
	\frac{h_6'' }{h_6 }
	-\frac{4}{r^2}+\frac{4 a  \omega  }{r^2}-\frac{2 a' }{r a }-\frac{2
	\omega ' }{r \omega  }+\frac{6 h_4' }{r h_4 }+\frac{\omega '  h_4' }{\omega   h_4 }-\frac{h_4' {}^2}{4 h_4 {}^2}-\frac{h_6' }{r
	h_6 }-\frac{a'  h_6' }{2 a  h_6 }-\frac{\omega '  h_6' }{2 \omega   h_6 }+\frac{3 h_4'  h_6' }{2 h_4  h_6 }-\frac{5 h_6' {}^2}{4
	h_6 {}^2}
	=0
\label{einstein_eq_mu2_h6}
\end{equation}
\begin{equation}
	\frac{\omega '' }{\omega  }
	-\frac{20}{r^2}+\frac{36 a }{r^2}-\frac{16 a  \omega  }{r^2}+\frac{\omega ' }{r \omega  }-\frac{a'  \omega ' }{2 a  \omega  }-\frac{\omega
	' {}^2}{2 \omega  ^2}+\frac{10 h_4' }{r h_4 }-\frac{h_4' {}^2}{4 h_4 {}^2}-\frac{10 h_6' }{r h_6 }+\frac{5 h_4'  h_6' }{2 h_4 
	h_6 }-\frac{5 h_6' {}^2}{4 h_6 {}^2}
	=0
\label{einstein_eq_mu2_omega}
\end{equation}
\begin{equation}
	\frac{20}{r^2}+\frac{4 \omega ' }{r \omega  }-\frac{10 h_4' }{r h_4 }-\frac{\omega '  h_4' }{\omega
	  h_4 }+\frac{h_4' {}^2}{4 h_4 {}^2}+\frac{10 h_6' }{r h_6 }+\frac{\omega '  h_6' }{\omega   h_6 }-\frac{5 h_4'  h_6' }{2
	h_4  h_6 }+\frac{5 h_6' {}^2}{4 h_6 {}^2}
	=
	a \left\{ \frac{24}{r^2} - \frac{4 \omega  }{r^2} - \frac{36 \mu _2^2}{r^6 g_s \omega   h_4  h_6 } \right\}
\label{einstein_eq_mu2_a}
\end{equation}
\label{einstein_eq_mu2_full}
\end{subequations}
Each of the first three equations here, eq.~\eqref{einstein_eq_mu2_h4}-\eqref{einstein_eq_mu2_omega}, is a second order ordinary differential equation w.r.t. $h_4(r)$, $h_6(r)$ and $\omega(r)$ respectively.
The fourth equation, eq.~\eqref{einstein_eq_mu2_a}, is algebraic in $a(r)$ and does not contain any second derivatives of other warp factors.
In principle one could substitute $a(r)$ from eq.~\eqref{einstein_eq_mu2_a} into eq.~\eqref{einstein_eq_mu2_h4}-\eqref{einstein_eq_mu2_omega} and obtain a system of only three differential equations for the three warp factors $h_4(r)$, $h_6(r)$ and $\omega(r)$ only.
However, the resulting system would be more complicated in certain respects, and we choose to work with eqs.~\eqref{einstein_eq_mu2_full} instead.

\subsubsection{Warp factors at large $r$}

At large $r \gg |b|^{1/3}, \sqrt{\mu_2}$ we expect small back reaction to the metric.
The warp factors approach the unperturbed value 1 in this limit, and this allows one to solve the Einstein equations \eqref{einstein_eq_mu2_full} perturbatively.
This task was  carried out in \cite{NS}\footnote{In eq.~\eqref{fluxed_mu2_large-r_solution} and eq.~\eqref{fluxed_mu1_large-r_solution} (see below)  we corrected numerical coefficients in the results obtained in  \cite{NS}, see the Mathematica code \cite{git_conifold_einstein}. Moreover, because here we  concentrate on the limit $|b|^{1/3} \ll \sqrt{\mu_1}, \sqrt{\mu_2}$, the UV cutoff of the logs is now determined by $\mu_1$ and $\mu_2$ rather then
 $|b|^{1/3}$ used in \cite{NS} in the opposite limit  $|b|^{1/3} \gg\sqrt{\mu_1}, \sqrt{\mu_2}$.} .
In this regime, solution of eq.~\eqref{einstein_eq_mu2_full} reads
%
%
%
\begin{equation}
\begin{gathered}
h_4=1+ \frac{9}{g_s}\,\frac{\mu_2^2}{r^4}\,\log{\frac{r}{\sqrt{\mu_2}}} + \ldots, \\
 h_6= 1+ \frac{9}{g_s}\,\frac{\mu_2^2}{r^4}\,\log{\frac{r}{\sqrt{\mu_2}}} 
+\ldots, \\
 a =1+\ldots, \\
\omega = 1+\frac3{g_s}\,\frac{\mu_2^2}{r^4} +\ldots.
\end{gathered} 
\label{fluxed_mu2_large-r_solution}
\end{equation}
Dots in eq.~\eqref{fluxed_mu2_large-r_solution} stand for sub-leading corrections of order of $O(\mu_2^4/r^8)$.

\subsubsection{Solving non-linear Einstein equations}
\label{sec:mu2_small-r}

When $r$ becomes small, of the order of $\sqrt{\mu_1}, \sqrt{\mu_2}$ or less, corrections in the approximate solutions eq.~\eqref{fluxed_mu2_large-r_solution} 
become important, and the perturbation theory breaks down.
In this regime one has to solve the Einstein equations exactly.
However, the systems of equations in \eqref{einstein_eq_mu2_full} present a challenge for finding a full analytical solution.
Our approach here is to solve these equations numerically and then match the large-$r$ behavior with the small-$r$ asymptotics (to be computed below).

To find the numerical solution we used the following strategy.
Because the behavior of the warp factors is known at large $r$ (see eq.~\eqref{fluxed_mu2_large-r_solution}), but unknown for small $r$, we  formulated the problem as a Cauchy problem.
From \eqref{fluxed_mu2_large-r_solution} we can infer the initial conditions for the warp factors and their first derivatives at a certain large value of $r$.
In the computation we started at $r = 10^3\sqrt{\mu_2}$, so that the next-to-leading order correction in \eqref{fluxed_mu2_large-r_solution} has a magnitude relative to the known, leading order correction, of the order $O(\mu^4/r^8) \sim 10^{-24}$.
We expect this to be a negligible correction.
In order to better capture the large-$r$ behavior within the machine accuracy, for the numerical algorithm we did an exponential substitution, $h_i(r) = e^{f_i(r)}$ and solved for $f_i(r)$, converting back to $h_i(r)$ at the very end; here, the index runs $i=4,6,a,\omega$ for the warp factors $h_4$, $h_6$, $h_a \equiv a$ and $h_\omega \equiv \omega$ respectively.
The solution is then propagated from large $r$ down to small $r$ by the Runge–Kutta 4$^\text{th}$ order method with non-uniform step size.
In our computation the solution was propagated from $r = 10^3\sqrt{\mu_2}$ down to $r = 10^{-2}\sqrt{\mu_2}$.
The step size was varied non-linearly from $\Delta r = 1\cdot \sqrt{\mu_2}$ at large $r$ where the solution is nearly constant, to $\Delta r = 10^{-4} \cdot \sqrt{\mu_2}$ at small $r$ where some of the warp factors vary more rapidly.

\begin{figure}[h]
    \centering
    \begin{subfigure}[t]{0.49\textwidth}
        \centering
        \includegraphics[width=\textwidth]{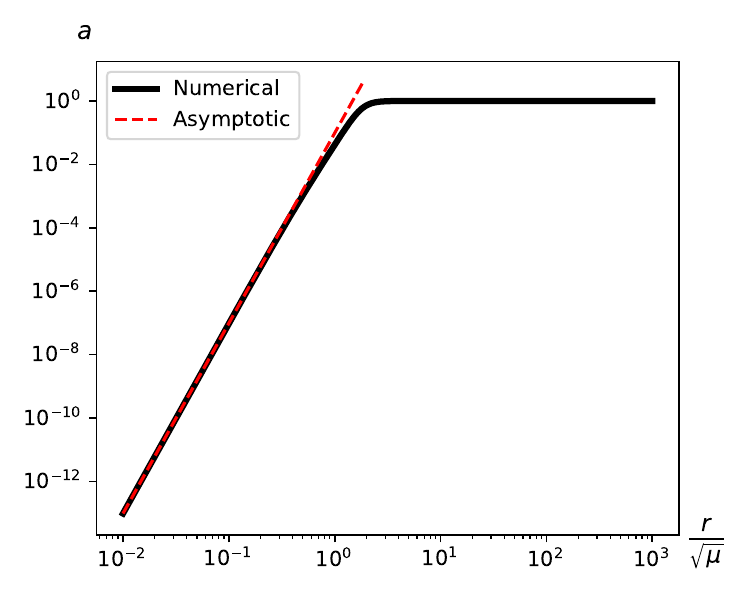}
        \label{fig:warp_factors_mu2_numerical_a}
    \end{subfigure}%
    ~ 
    \begin{subfigure}[t]{0.49\textwidth}
        \centering
        \includegraphics[width=\textwidth]{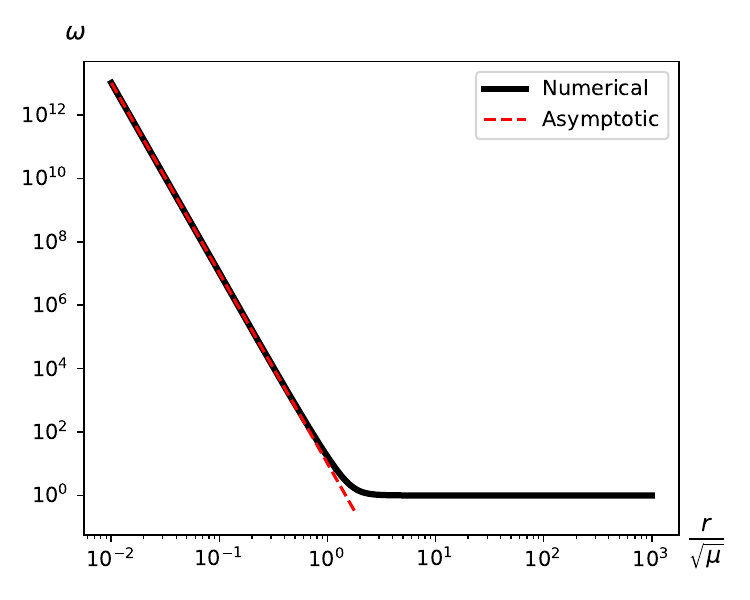}
        \label{fig:warp_factors_mu2_numerical_w}
    \end{subfigure}%
    
    \begin{subfigure}[t]{0.49\textwidth}
        \centering
        \includegraphics[width=\textwidth]{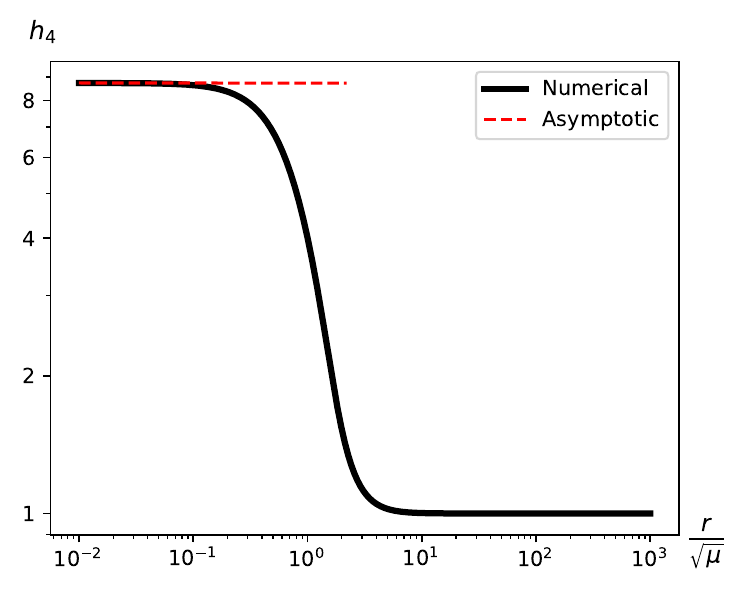}
        \label{fig:warp_factors_mu2_numerical_h4}
    \end{subfigure}%
    ~ 
    \begin{subfigure}[t]{0.49\textwidth}
        \centering
        \includegraphics[width=\textwidth]{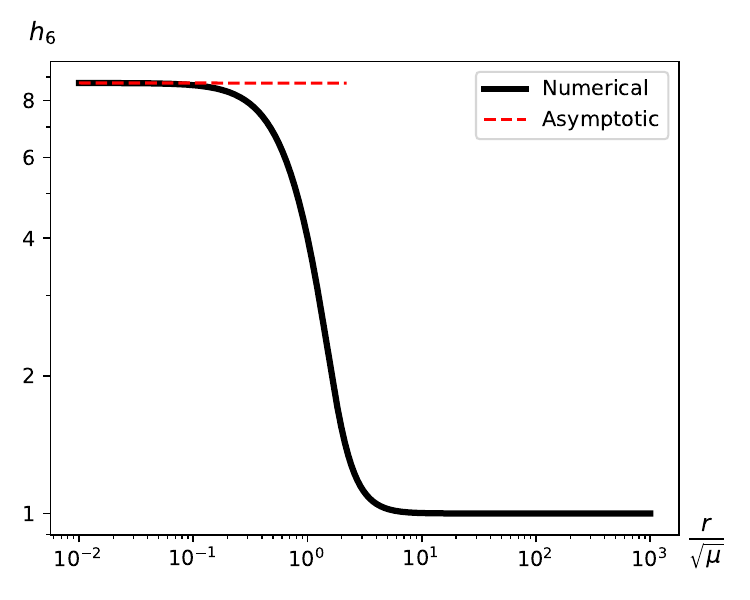}
        \label{fig:warp_factors_mu2_numerical_h6}
    \end{subfigure}%
	\caption{
		Warp factors $a$, $\omega$, $h_4$ and $h_6$ as functions of $r$. 
		Black solid lines are the numerical solution, red dashed lines represent the small-$r$ asymptotic eq.~\eqref{warp_factors_small-r_power} with eq.~\eqref{warp_factors_mu2_numerical_coefficients}
	}
\label{fig:warp_factors_mu2_numerical}
\end{figure}

The result of our numerical calculation is shown on Fig.~\ref{fig:warp_factors_mu2_numerical}.
While for large $r$ the solution follows the asymptotic \eqref{fluxed_mu2_large-r_solution}, around $r \sim \sqrt{\mu_2}$ the behavior of the warp factors change to the power-like,
\begin{equation}
	 h_i \approx C_i  r^{\alpha_i},
\label{warp_factors_small-r_power}
\end{equation}
where  $C_i$, $\alpha_i$ are constants.  The scale of coefficients $C_i$ is determined by $\mu_2$, 
\beq
C_i \sim \mu_2^{-\frac{\alpha_i}{2}}. 
\eeq
Specifically, for this solution we find the following values of the coefficients (setting $\mu_2=1$ for simplicity):
\begin{equation}
\begin{aligned}
	\alpha_4 \approx \alpha_6 &\approx 0.000 \,, \quad 	& C_4 &\approx C_6 \approx 8.727 \,, \\
	\alpha_a &\approx 6.000 \,, \quad 					& C_a &\approx 0.09700 \,, \\
	\alpha_\omega &\approx -6.000 \,, \quad 			& C_\omega &\approx 10.307 \,.
\end{aligned}
\label{warp_factors_mu2_numerical_coefficients}
\end{equation}
Note that
\begin{equation}
    \alpha_a + \alpha_\omega \approx 0.000,   \quad   C_a  C_\omega \approx 1.000.
\end{equation}
We checked that the asymptotic eq.~\eqref{warp_factors_small-r_power} with eq.~\eqref{warp_factors_mu2_numerical_coefficients} satisfies (to the leading order in $r$) the Einstein equations \eqref{einstein_eq_mu2_full}.
One thing to note here is that the r.h.s. of eq.~\eqref{einstein_eq_mu2_h4}-\eqref{einstein_eq_mu2_omega} becomes small as $r$ tends to zero,
\begin{equation}
	\frac{a }{r^6 g_s \omega   h_4  h_6 } 
	\sim r^{\alpha_a - \alpha_{h4} - \alpha_{h6} - \alpha_\omega - 6 } 
	\approx r^6
\end{equation}
This can be contrasted to a typical term on the l.h.s. which grow as $\sim 1/r^2$ at small $r$ (see also eq.~\eqref{powers_eq_main_mu2} below).
We  conclude that to the leading order at small $r$ the $\mu_2$-dependent term in eq.~\eqref{einstein_eq_mu2_full} is in fact negligible.

To understand our results better, let us compute the small-$r$ asymptotic analytically.
Here we will consider one particular possibility of the small-$r$ behavior that is supported by the numerical solution above.

Suppose that the warp factors at small $r$ indeed follow the asymptotic \eqref{warp_factors_small-r_power} with some coefficients to be determined from the Einstein equations.
Guided by the numerical results, we also suppose that these coefficients are such that the r.h.s. of eq.~\eqref{einstein_eq_mu2_full} can be neglected to the leading order.
Then, substituting the ansatz \eqref{warp_factors_small-r_power} into the equations \eqref{einstein_eq_mu2_full}, after a bit of algebra we arrive at the following system: 

\begin{subequations}
\begin{multline}
	-16 r^{\alpha _a} C_a \left(-6+r^{\alpha _{\omega }} C_{\omega }\right)-5 \alpha _4^2+2 \alpha _4 \left(28+7
	\alpha _6-\alpha _a+3 \alpha _{\omega }\right) \\
	-\left(4+\alpha _6\right) \left(20+5 \alpha _6+4 \alpha _{\omega }\right)
	=0
\label{powers_eq_main_mu2_4}
\end{multline}
\begin{multline}
	16 r^{\alpha
	_a+\alpha _{\omega }} C_a C_{\omega }-\alpha _4^2+\alpha _4 \left(6 \alpha _6+4 \left(6+\alpha _{\omega }\right)\right)-\left(4+\alpha
	_6\right) \left(\alpha _6+2 \left(2+\alpha _a+\alpha _{\omega }\right)\right) 
	=0
\label{powers_eq_main_mu2_6}
\end{multline}
\begin{multline}
	-80-16 r^{\alpha _a} C_a \left(-9+4
	r^{\alpha _{\omega }} C_{\omega }\right)-\alpha _4^2-40 \alpha _6-5 \alpha _6^2+10 \alpha _4 \left(4+\alpha _6\right)-2 \alpha _a \alpha
	_{\omega }+2 \alpha _{\omega }^2 
	=0
\label{powers_eq_main_mu2_omega}
\end{multline}
\begin{multline}
	\alpha _4^2-2 \alpha _4 \left(20+5 \alpha _6+2 \alpha _{\omega }\right)+\left(4+\alpha
	_6\right) \left(20+5 \alpha _6+4 \alpha _{\omega }\right) \\
	+16 r^{\alpha _a} C_a \left(-6+r^{\alpha _{\omega }} C_{\omega }+\frac{9
	r^{-4-\alpha _4-\alpha _6-\alpha _{\omega }} \mu _2^2}{C_4 C_6 C_{\omega } g_s}\right)
	=0
\label{powers_eq_main_mu2_a}
\end{multline}
\label{powers_eq_main_mu2}
\end{subequations}
We want to study the limit $r \to 0$.
One can see that eq.~\eqref{powers_eq_main_mu2} has terms with powers of $r$, and if these powers are positive, we can neglect these terms at small $r$.
Numerical results suggest that we need to look for a solution satisfying
\begin{equation}
	\alpha_a > 0 \,, \quad
	\alpha _a + \alpha _{\omega } = 0 \,, \quad
	-4-\alpha _4-\alpha _6+\alpha_a-\alpha _{\omega } > 0
\label{power_constraints_1}
\end{equation}
Assuming these constrains, the system eq.~\eqref{powers_eq_main_mu2} in the limit of small $r$ becomes a system of algebraic equations.
The only solution with all the warp factors non-zero is the following:
\begin{equation}
	\alpha_a = \frac{3}{2} (4 + \alpha_6) \,, \quad
	\alpha_\omega = - \frac{3}{2} (4 + \alpha_6) \,, \quad
	\alpha_4 = 0 \,, \quad
	C_a C_{\omega } = \frac{1}{16} (4 + \alpha_6)^2
\label{warp_factors_mu2_exact_coefficients}
\end{equation}
The power $\alpha_6$ and the coefficients $C_4$, $C_6$ are arbitrary.
One can see that the numerical results in eq.~\eqref{warp_factors_mu2_numerical_coefficients} precisely correspond to the solution in eq.~\eqref{warp_factors_mu2_exact_coefficients} with $\alpha_6 = 0$:
\begin{equation}
	\alpha_a = 6 \,, \quad
	\alpha_\omega = - 6 \,, \quad
	\alpha_4 = 0 \,, \quad
	\alpha_6 = 0 \,, \quad
	C_a C_{\omega } = 1
\label{warp_factors_mu2_exact_coefficients_ourcase}
\end{equation}

Here we would like pause to make a brief remark.
Note that the solution \eqref{warp_factors_mu2_exact_coefficients} depends on an arbitrary power coefficient $\alpha_6$.
The source of this arbitrariness is the reparametrization invariance.
The idea is that any solution of the Einstein equations should transform covariantly under the change of coordinates on the conifold.
In particular, for the power-like  behavior in \eqref{warp_factors_small-r_power} at small $r$ we can redefine 
 the radial coordinate, $r = \hat{r}^\beta$ with some non-zero constant $\beta$.
Under this change of variables we have $r^2 = \hat{r}^2 \hat{r}^{2\beta-2}$ and $d r^2 = \beta^2 \hat{r}^{2\beta-2} d\hat{r}^2$. Then in terms of the new radial coordinate $ \hat{r}$ the metric of the internal 6D space takes the same form
\eqref{warpedconi} with warp factors
 $h_i = \hat{C}_i \hat{r}^{\hat{\alpha}_i}$, where the coefficients $\hat{C}_i$, $\hat{\alpha}_i$ are given by
\begin{subequations}
\begin{equation}
	\hat{\alpha}_4 = \beta \alpha_4 \,, \quad
	\hat{\alpha}_6 + 4 = \beta ( \alpha_6 + 4 ) \,, \quad
	\hat{\alpha}_a = \beta \alpha_a \,, \quad
	\hat{\alpha}_\omega = \beta \alpha_\omega
\end{equation}
\begin{equation}
	\hat{C}_a = \beta^2 C_a \,, \qquad
	\hat{C}_{i}  = C_{i}, \qquad  i=4,6,\omega
\end{equation}
\label{reparametrization_transformation}
\end{subequations}
One can see that  the solution \eqref{warp_factors_mu2_exact_coefficients} transforms consistently under reparametrization \eqref{reparametrization_transformation}. We can use this reparametrization covariance to fix the value
of  $\alpha_6$  . Namely, starting with arbitrary $\hat{\alpha}_6$ we fix the gauge requiring that $\alpha_6 =0$. In this gauge the solution  \eqref{warp_factors_mu2_exact_coefficients} reduces to  \eqref{warp_factors_mu2_exact_coefficients_ourcase}. Note that finding the large-$r$ asymptotics 
\eqref{fluxed_mu2_large-r_solution}, which we used as initial conditions  also requires fixing the  gauge \cite{NS}.
Of course, these choices of gauge in  large-$r$ and small-$r$ solutions have to match each other.
It turns out that in our numerical algorithm  starting with a particular form of the solution at large $r$  fixes unambiguously the small-$r$ behavior  given by eq.~\eqref{warp_factors_mu2_exact_coefficients_ourcase}.

One might worry that, in the limit $\mu_2 \to 0$, the solution that we found for warp factors does not seem to go smoothly to the metric of unwarped conifold compactification on $\mathbb{R}^{1,3} \times Y_6$.
However this is not the case.
When $\mu_2$ becomes sufficiently small such that $|b| \gg \mu_2^{3/2}$, we return to the situation studied in Sec.~\ref{sec:def}.
Effectively, the transition point of the solution at $r \sim \sqrt{\mu_2}$ discussed above now goes under the cutoff at $r \sim |b|^{1/3}$ and is not seen.
After that it becomes evident that the limit $\mu_2 \to 0$ is indeed smooth. In particular,  the UV cutoff of logarithms of the large $r$ asymptotics \eqref{fluxed_mu2_large-r_solution} in the small $\mu_2$ limit is replaced  by $|b|^{1/3}$.

\subsubsection{Baryon potential}

Let us reiterate what we know.
We have the large-$r$ asymptotic of the warp factors, eq.~\eqref{fluxed_mu2_large-r_solution}.
We also have the small-$r$ asymptotic, eq.~\eqref{warp_factors_mu2_exact_coefficients_ourcase}.
Numerical solution tells us that these asymptotics indeed correspond to the same solution.
Thus we are in a position to compute the potential for the field $b$.

To derive the potential generated by the $H_3$ flux, we need to substitute our solutions for the metric and for the $H_3$ form into the 10D action.
To simplify the computation we can again use the formula \eqref{pot_gen}, because it is covariant and suitable for any background:
\begin{equation}
	V(b)=\frac{T^4}{(2\pi)^3g_s^2}\,\int d^6 x \sqrt{g_6}\;\frac{e^{-\Phi}}{24}\, (H_3)^2 
\label{pot_gen_3}
\end{equation}
The $H_3$ form and it's dual are given in eq.~\eqref{H3_second_solution_warped} and eq.~\eqref{H3_second_solution_warped_dual}.
One can compute the integrand of eq.~\eqref{pot_gen_3} directly in tensor notation.
We have
\begin{equation}
	(H_3)^2 \equiv H_{MNL} H^{MNL} = 2^4 3^3   \frac{ \mu_2^2}{r^6 \omega h_6^{3/2}}
\end{equation}
The dilaton gives $e^{-\Phi}=1/(g_s h_4)$, see eq.~\eqref{dilaton_solution}.
Integrating this with the measure \eqref{Vol_warped} we arrive at
\begin{equation}
	V(b) = \frac{4}{3} \frac{T^4 \mu^2_2  }{g_s^3} \int \frac{dr}{r}  \frac{a^{1/2}}{h_4^{2}\, \omega^{1/2}} 
\label{potential_warped_mu2_1}
\end{equation}
Alternatively, one can do the computation as an integral of the form $H_3 \wedge (e^{-\Phi} \ast H_3)$ using eq.~\eqref{H3_second_solution_warped} and eq.~\eqref{H3_second_solution_warped_dual}, the result is the same.

Now, recall that in Sec.~\ref{sec:def} we computed the potential $V(b)$ neglecting the back reaction of the $H_3$ flux on the metric, i.e. in the limit $|b|^{2/3} \gg \mu_2$.
The resulting potential (see eq.~\eqref{Veff_noback reaction_full} and Fig.~\ref{fig:Veff_noback reaction}) pushes $b$ towards large values and leads to a runaway vacuum.
In this Section we want to check whether the inclusion of the back reaction effect changes  this behavior.

To this end let us analyze the integral in \eqref{potential_warped_mu2_1}.
We know the behavior of the warp factors in the limit $|b|^{2/3} \ll \mu_2$ with the back reaction accounted for.
We can split the integration domain into two regions, $r < \sqrt{\mu_2}$ and $r > \sqrt{\mu_2}$.

In second region, $r > \sqrt{\mu_2}$, the warp factors actually do not depend on $b$, see eq.~\eqref{fluxed_mu2_large-r_solution}.
Therefore, the integral in eq.~\eqref{potential_warped_mu2_1} over this region only gives some constant of the order of $\sim \log(R_\text{IR}/\sqrt{\mu_2})$, where $R_\text{IR}$ is the IR cutoff of the conifold radial coordinate $r$.

The first region, $r < \sqrt{\mu_2}$, in principle could give a non-trivial $b$-dependence, if the integral in eq.~\eqref{potential_warped_mu2_1} depended on the details of the UV cutoff region at $r \to 0$.
Assuming the power-like behavior of the warp factors, eq.~\eqref{warp_factors_small-r_power}, we obtain for this integral
\begin{equation}
	V(b) \approx \frac{4}{3} \frac{T^2}{g_s^3} \frac{C_a^{1/2}}{C_\omega^{1/2} C_{h4}^2} \mu_2^2 
\int_{|b|^{1/3}}^{\sqrt{\mu_2}} \frac{dr}{r} r^\gamma \,, \quad
	\gamma = \frac{\alpha_a}{2}-\frac{\alpha_\omega}{2}-2\alpha_{4}
\label{potential_warped_mu2_2}
\end{equation}
In our case  eq.~\eqref{warp_factors_mu2_exact_coefficients_ourcase} gives $\gamma = 6$, so that the integral \eqref{potential_warped_mu2_2} is convergent at $r \to 0$.
Therefore, it does not depend on the details of the conifold deformation by $b$ in the limit $|b |\ll \mu_2^{3/2}$.

\begin{figure}[h]
    \centering
    \begin{subfigure}[t]{0.49\textwidth}
        \centering
        \includegraphics[width=\textwidth]{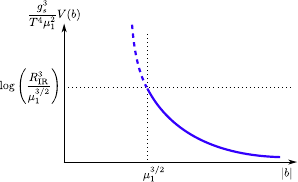}
        \caption{$\mu_1$ switched on, $\mu_2=0$}
        \label{fig:proposed_potential_full_mu1}
    \end{subfigure}%
    ~ 
    \begin{subfigure}[t]{0.49\textwidth}
        \centering
        \includegraphics[width=\textwidth]{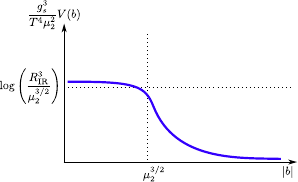}
        \caption{$\mu_1=0$, $\mu_2$ switched on}
        \label{fig:proposed_potential_full_mu2}
    \end{subfigure}%
	\caption{
		The  schematic shape of the baryon potential $V(b)$ generated by the $H_3$ form flux after including the back reaction effects.
		In both cases the potential drives $b$ to large values and leads to a runaway vacuum.
	}
\label{fig:proposed_potential_full}
\end{figure}

We  conclude that the potential has the following shape.
For a given flux parameter $\mu_2$, the potential is flat at small $|b |\ll \mu_2^{3/2}$ and falls off at large $|b| \gg \mu_2^{3/2}$ as discussed in Sec.~\ref{sec:nobackreaction_Vb}, see Fig.~\ref{fig:proposed_potential_full_mu2}.
Thus we indeed have a runaway vacuum for $b$.

\subsection{Baryon potential from $\mu_1$ flux}
\label{potential_from_mu1}

Now we come to discussing the first solution for the $H_3$ form parametrized by $\mu_1$, eq.~\eqref{H3_first_solution_warped}.
This case turns out to be less tractable.

\subsubsection{Einstein equations}

Now let us turn to the first solution for the $H_3$ form parametrized by $\mu_1$, see eq.~\eqref{H3_first_solution_warped}.
The corresponding Einstein equations are derived in Appendix~\ref{sec:warped_sing_ricci}, see 
  equations \eqref{ricci_einstein_eq_mu1}, where components of Ricci tensor is given by eqs.~\eqref{l.h.s.1}-\eqref{l.h.s.4}.
Since the potential $V(b)$ is known in the limit $|b| \gg \mu_1^{3/2}$ (see eq.~\eqref{Veff_noback reaction_full}), here we  concentrate on the opposite limit $|b| \ll \mu_1^{3/2}$.

\subsubsection{Warp factors at large $r$}

At large $r \gg |b|^{1/3}, \sqrt{\mu_1}$ we expect small back reaction to the metric.
The warp factors approach unity in this limit, and this allows one to solve the Einstein equations \eqref{ricci_einstein_eq_mu1} perturbatively.
This  done in \cite{NS}.
For the $\mu_1$-flux, solution of eq.~\eqref{ricci_einstein_eq_mu1} reads
%
\begin{equation}
\begin{gathered}
h_4=1+ \frac{9}{g_s}\,\frac{\mu_1^2}{r^4}\,\log{\frac{r}{\sqrt{\mu_1}}} + \ldots, \\
 h_6= 1+ \frac{9}{g_s}\,\frac{\mu_1^2}{r^4}\,\log{\frac{r}{\sqrt{\mu_1}}} 
+\ldots, \\
 a =1+\ldots, \\
\omega = 1-\frac3{g_s}\,\frac{\mu_1^2}{r^4} +\ldots.
\end{gathered} 
\label{fluxed_mu1_large-r_solution}
\end{equation}
Dots in eq.~\eqref{fluxed_mu1_large-r_solution} stand for sub-leading corrections of order of $O(\mu_1^4/r^8)$.

\subsubsection{Baryon potential}

As in the previous case we attempted to follow the same strategy as in Sec.~\ref{sec:mu2_small-r}, namely to solve  equations \eqref{ricci_einstein_eq_mu1} numerically and then infer the small-$r$ asymptotic analytically.

In the numerical algorithm we have set Cauchy initial conditions (inferred from eq.~\eqref{fluxed_mu1_large-r_solution}) and propagated them down to small $r$ using the Einstein equations \eqref{ricci_einstein_eq_mu1}. 
However, contrary to what has been found for the $\mu_2$ flux in Sec.~\ref{sec:mu2_small-r}, in the present case with the $\mu_1$ flux the numerical algorithm did not produce a solution that would go smoothly to small $r$.
Instead, the numerical solution blows up near $r \sim \sqrt{\mu_1}$, and the algorithm breaks down.

This suggests that, in fact, the exact solution of the Einstein equations \eqref{ricci_einstein_eq_mu1} has a naked singularity at a finite value of $r =r_0\sim \sqrt{\mu_1}$ where curvature becomes infinite. 
 (see Appendix~\ref{sec:warped_sing_ricci} for the relevant formulas).
Although we have not been able to reliably establish the existence of such a singularity, we have a strong hint of its presence.
Let us consider possible implications of that.

There must be a mechanism to smoothen this singularity (see also \cite{Johnson:1999qt} for a related discussion).
 In our case it is natural to expect a generation of a potential $V(b)$ that would push $b$ towards large values, $|b|^{1/3} \gg \sqrt{\mu_1}$.

Therefore, we can suggest that the potential $V(b)$ has the form as shown on Fig.~\ref{fig:proposed_potential_full_mu1}.
Since the geometry is defined only at $r>r_0$ the potential does not depend on $b$ at $|b|\ll r_0^{3}$. At 
 $|b| \gtrsim r_0^3$  this potential is strongly repulsive.
For larger $b$,  $|b| \gg \mu_1^{3/2}$ the potential is given by eq.~\eqref{Veff_noback reaction_full}; it is again a repulsive potential leading to a runaway vacuum for $b$.

\section {Implications for 4D SQCD }
\label{sec:quarkmasses}

In this Section we interpret the above results in terms of the 4D SQCD that we briefly reviewed in  Sec.~\ref{sec:NAstring}.
We will largely follow \cite{Y_NSflux,NS} with a few additional remarks.

\subsection{3-form flux in terms of quark masses}

In \cite{Y_NSflux,NS} the flux compactification with the $H_3$ form was interpreted as switching on a particular combination of quark masses.
The basic idea is as follows.

In terms of the field content of \ntwo SQCD, the only scalar potential deformation allowed by \ntwo supersymmetry is the mass term for quarks\footnote{Note that quarks are perturbative states with flat K\"ahler potential. For a field with a non-trivial K\"ahler potential   it is possible to have a non-trivial scalar potential without breaking of  \ntwo supersymmetry \cite{Louis2}. For example, the field $b$ is  of this type.}.
In general, for $N_f=4$ flavors one can introduce four mass parameters $m_i$, $i = 1,\ldots,4$.
Because we can make a shift of the complex scalar of the \ntwo U(1) vector multiplet, only three parameters are independent, say,
\begin{equation}
	m_1-m_2, \qquad m_3-m_4, \qquad m_1-m_3
\label{massdifferences}
\end{equation}
Since the $b$-baryon mass \eqref{m_b} in not generated (see Sec.~\ref{sec:nobackreaction_Vb}), we have a constraint
\beq
m_1 +m_2 -m_3 -m_4 =0.
\label{constraint1}
\eeq
Another constraint  is
\beq
m_1 m_2-m_3 m_4 =0.
\label{constraint2}
\eeq
It is imposed to avoid infinite VEV of $\sigma$ (a scalar superpartner of the U(1) gauge field), which would costs an infinite energy in the world sheet  \wcpt model at large $b$, see \cite{Y_NSflux} for details.
After imposing these constrains we can see that the $H_3$-flux can be interpreted in terms of a single mass difference $(m_1-m_2)$. 
 We define a complex parameter  $\mu $ and  identify  \cite{Y_NSflux}
\beq 
\mu \equiv \mu_1 + i\mu_2 = {\rm const}\,\sqrt{\frac{g_s^3}{T}}\, (m_1-m_2), \qquad m_3=m_1, \qquad m_4=m_2.
\label{mum}
\eeq

Note that, in fact, the potential $V(b)$ depends on the combination of the flux parameters given by 
\begin{equation}
    (\mu_1^2 +\mu_2^2) = |\mu|^2 = {\rm const}\,\frac{g_s^3}{T}\, |m_1-m_2|^2.
\label{mu_quadratic_combination}
\end{equation}
To see this  note, that
at $|b|  \gtrsim\mu_i^{3/2}$, $i=1,2$ the leading contribution to the potential $V(b)$
is given by the logarithmic term in \eqref{pot_log} \footnote{Strictly speaking, this statement is valid only at $|b|$ not too close to $r_0^3$,  $r_0\sim \sqrt{\mu_1}$. 
 As was discussed in Sec.~\ref{potential_from_mu1}, it is possible that at  $|b|\to r_0^3$ there is a  naked singularity of the solution with the $\mu_1$ flux parameter.  In this limit the supergravity approximation is no longer applicable, and we cannot reliably determine the potential by this method.} . Taking the warp factors into account in  \eqref{potential_warped_mu2_1} or similar formula for the potential induced by $\mu_1$-flux gives additional
contribution given by a finite function of rations   $\mu_1/|b|^{2/3}$ and $\mu_2/|b|^{2/3}$, which can be neglected as compared to the infrared logarithmic term in the limit $R_\text{IR}\to\infty$. Moreover, in the limit of very large $b$, $|b| \to b_\text{IR}$ corrections in powers of 
$\mu_i^2/|b|^{4/3}$ becomes small and warp factors tend to unity \cite{NS}.  In this limit the potential is given by eq.~\eqref{pot_b_to_b_IR}. 

Now let us summarize the impact of nonzero quark mass difference $(m_1-m_2)$ in 4D SQCD.
In particular, if we take  the limit  of large $|m_1-m_2| \gg \sqrt{\xi}$ in 4D SQCD keeping the constraints  $m_3=m_1$ and $m_4=m_2$  off-diagonal gauge fields together with two quark flavors acquire large masses $\sim |m_1-m_2|$ and decouple. As
non-Abelian degrees of freedom decouple the  U(2) gauge theory
flows to two non-interacting \ntwo supersymmetric QED with the gauge groups U(1) and $N_f=2$ quark flavors in each. 
In this case   the \wcpt model on the non-Abelian string flows to \wcpo.
This means that the non-Abelian string  flows to an Abelian one \cite{Y_NSflux,NS}.

On the string theory side switching on $(m_1-m_2)$ is reflected in the degeneration of the conifold, which effectively reduces its dimension. Also in the limit 
$|b|\to\infty$ the radius of the sphere $S_3$ of the conifold  becomes infinite and it tends to a flat three dimensional space. This matches the field theory result \cite{AharSeib} that \wcpo model flows to a free theory in the infrared. It would be tempting to interpret the extra coordinate of the sphere $S_3$ of the conifold in the limit $|b|\to\infty$ as a Liouville coordinate for a non-critical string associated with the  \wcpo model. This is left for a future work.

\section{Conclusions}
\label{sec:concl}

In this work we analyzed a deformation of the string theory for the critical non-Abelian vortex in \ntwo SQCD with gauge group U(2) and $N_f=4$ quark flavors, incorporating an NS 3-form flux, building on results from the previous works \cite{Y_NSflux,NS}. 
Using a supergravity approach, we derived an exact solution for the 3-form $H_3$ in the fixed background of the deformed conifold.
The presence of the 3-form $H_3$ generates a potential for the conifold’s complex structure modulus $b$, which corresponds to a massless BPS baryonic hypermultiplet in 4D SQCD at strong coupling. 
This potential lifts the Higgs branch formed by the VEVs of $b$, resulting in a runaway vacuum for $b$, where $\langle b \rangle \to \infty$. 

In the second part of this paper we included the back reaction on the conifold metric.
The metric is significantly modified for small $b$.
However, we find that the vacuum is still of the runaway type.

At the runaway vacuum, the conifold effectively reduces to lower dimensions. 
This aligns qualitatively with a flow to the \wcpo model on the string world sheet, expected upon introduction of a certain mass combination, decoupling one $n$-field and one $\rho$-field. 
In 4D SQCD, this corresponds to a flow toward \ntwo supersymmetric QED with two charged flavors.

\section*{Acknowledgments}

The work of P.P. and A.Y. is supported by the Foundation for the
Advancement of Theoretical Physics and Mathematics \textquote{BASIS}, Grant No.
22-1-1-16. The work of E.I. is supported by U.S. Department of Energy Grant No. de-sc0011842.

\appendix

\section{Results for the Christoffel symbols and the Ricci tensor}
\label{sec:warped_sing_ricci}

In this Appendix we provide the expressions for the Christoffel symbols and the Ricci tensor computed for the warped metric \eqref{warped_singular_coni_full_metric}.
This computation was performed in Wolfram Mathematica \cite{git_conifold_einstein}.

Christoffel symbols for the warped metric \eqref{warped_singular_coni_full_metric} organize into blocks proportional to the original conifold metric in eq.~\eqref{10metric}-\eqref{angles}.
We denote the latter as $g^{(c)}$.
The non-zero components of the Christoffel symbols are as follows:
\begin{equation}
\begin{aligned}
	\Gamma^{r}_{\alpha\beta} &= -\frac{g^{(c)}_{\alpha\beta}}{a}\left( \frac{1}{r } + \frac{h_6'}{4  h_6} \right), \\
	\Gamma^{\beta}_{r\alpha} &= \Gamma^{\beta}_{\alpha r}=\delta^{\beta}_{\alpha} 
		\left( \frac{1}{r}+\frac{h_6'}{4	h_6} \right), \\
	\Gamma^{r}_{\psi\psi} &= g^{(c)}_{\psi \psi} \left( -\frac{\omega }{r a}-\frac{\omega '}{2 a}-\frac{\omega  h_6'}{4 a h_6} \right), \\
	\Gamma^{\psi}_{r\psi} &= \Gamma^{\psi}_{\psi r} = \frac{1}{r}+\frac{\omega '}{2 \omega }+\frac{h_6'}{4 h_6}, \\
	\Gamma^{r}_{r r} &= \frac{a'}{2 a}+\frac{h_6'}{4 h_6}.
\end{aligned}
\end{equation}
Here, $\alpha , \beta$ run over $e_{\varphi_1}, e_{\varphi_2}, e_{\theta_1}, e_{\theta_2}$, while the index $\psi$ denotes the component corresponding to $e_{\psi}$.

Using Christoffel symbols we can find the Ricci tensor:
\begin{equation}\label{l.h.s.1}
	a {h_4}^{1/2} {h_6}^{1/2} R_{\mu \nu} = \eta_{\mu \nu} \left( 
		\frac{5 h_4'}{4 r h_4{} }-\frac{a' h_4'}{8 a h_4{} }+\frac{\omega
			' h_4'}{8  \omega  h_4{} }-\frac{h_4'{}^2}{2  h_4{}^{3} }+\frac{h_4' h_6'}{4 
			h_4{} h_6}+\frac{h_4''}{4  h_4{}}
	 \right),		
\end{equation}
\begin{equation}\label{l.h.s.2}
	a R_{\alpha\beta} = g^{(c)}_{\alpha\beta}  \left(
		\frac{6 a}{r^2 }-\frac{4}{r^2 }-\frac{2 a \omega}{r^2}+\frac{a'}{2 r a}-\frac{\omega '}{2 r  \omega }+\frac{h_4'}{r  h_4}-\frac{9 h_6'}{4 r  h_6}+\frac{a'
			h_6'}{8 a h_6}-\frac{\omega ' h_6'}{8  \omega  h_6}+\frac{h_4' h_6'}{4  h_4 h_6}-\frac{h_6''}{4 
			h_6}
	\right),
\end{equation}
\begin{equation}\label{l.h.s.3}
\begin{gathered}
\frac{a}{\omega} R_{\psi\psi} = g^{(c)}_{\psi\psi} \Big(-\frac{4}{r^2 }+\frac{4a \omega }{r^2}+\frac{  a'}{2 r a}-\frac{3 \omega '}{r \omega }+ 
\frac{a'\omega '}{4 \omega a}+\frac{\omega '^2}{4  \omega^2 }+
\frac{ h_4'}{r  h_4}+\frac{\omega ' h_4'}{2 \omega h_4}- \\
-\frac{9 h_6'}{4 r  h_6}+\frac{  a' h_6'}{8 a h_6}-
\frac{5 \omega ' h_6'}{8  \omega h_6}+\frac{h_4' h_6'}{4  h_4 h_6}-\frac{\omega ''}{2 \omega }-\frac{ h_6''}{4  h_6} \Big),
\end{gathered}
\end{equation}
\begin{equation}\label{l.h.s.4}
\begin{gathered}
	R_{rr} = \frac{5 a'}{2 r a}-\frac{\omega
		'}{r \omega }+\frac{a' \omega '}{4 a \omega }+\frac{\omega '^2}{4 \omega ^2}-\frac{a' h_4'}{2 a h_4}-\frac{5h_4'{}^2}{4 h_4{}^2}-\frac{5 h_6'}{4 r h_6}+\frac{5 a' h_6'}{8 a h_6}- \\
		- \frac{\omega ' h_6'}{8 \omega  h_6}-\frac{h_4'
		h_6'}{4 h_4 h_6}+\frac{5 h_6'{}^2}{4 h_6{}^2}-\frac{\omega ''}{2 \omega }+\frac{h_4''}{h_4}-\frac{5 h_6''}{4 h_6}.
\end{gathered} 
\end{equation}
Various overall factors are introduced here for convenience.

Using the expressions for the $H_3$ form, eq.~\eqref{H3_first_solution_warped} and \eqref{H3_second_solution_warped}, and the dilaton eq.~\eqref{dilaton_solution}, we can calculate the r.h.s. of the Einstein’s equations \eqref{Einstein}.
Taking the first expression for the $H_3$ form parametrized by $\mu_1$, eq.~\eqref{H3_first_solution_warped}, we obtain
\begin{subequations}
\begin{equation}\label{r.h.s.1st.1}
	a {h_4}^{1/2} {h_6}^{1/2} R_{\mu \nu} = -\eta_{\mu \nu} \frac {9 \mu_{1}^2}{g_s} \frac{h_4^3 a}{h_6 \omega} \frac{1}{r^6},		
\end{equation}
\begin{equation}\label{r.h.s.1st.2}
	a R_{\alpha\beta} = g^{(c)}_{\alpha\beta}  \frac {9\mu_{1}^2}{g_s} \frac{h_4^3 a}{h_6 \omega} \frac{1}{r^6},
\end{equation}
\begin{equation}\label{r.h.s.1st.3}
	\frac{a}{\omega} R_{\psi\psi} = -g^{(c)}_{\psi\psi}  \frac {9\mu_{1}^2}{g_s} \frac{h_4^3 a}{h_6 \omega} \frac{1}{r^6},
\end{equation}
\begin{equation}\label{r.h.s.1st.4}
	R_{rr} =  \frac {27\mu_{1}^2}{g_s} \frac{h_4^3 a}{h_6 \omega} \frac{1}{r^6}+\frac{1}{2}\frac{h_4 '^2}{h_4^2}.
\end{equation}
\label{ricci_einstein_eq_mu1}
\end{subequations}
For the second expression for the $H_3$ form parametrized by $\mu_2$, eq.~\eqref{H3_second_solution_warped}, we obtain
\begin{subequations}
\begin{equation}\label{r.h.s.2st.1}
a {h_4}^{1/2} {h_6}^{1/2} R_{\mu \nu} = -\eta_{\mu \nu} \frac {9 \mu_{2}^2}{g_s} \frac{ a}{h_6 h_4 \omega} \frac{1}{r^6},		
\end{equation}
\begin{equation}\label{r.h.s.2st.2}
	a R_{\alpha\beta} = g^{(c)}_{\alpha\beta}  \frac {9 \mu_{2}^2}{g_s} \frac{ a}{h_6 h_4 \omega} \frac{1}{r^6},
\end{equation}
\begin{equation}\label{r.h.s.2st.3}
	\frac{a}{\omega} R_{\psi\psi} = g^{(c)}_{\psi\psi}  \frac {27 \mu_{2}^2}{g_s} \frac{ a}{h_6 h_4 \omega} \frac{1}{r^6},
\end{equation}
\begin{equation}\label{r.h.s.2st.4}
	R_{rr} =  -\frac {9 \mu_{2}^2}{g_s} \frac{ a}{h_6 h_4 \omega} \frac{1}{r^6}  +   \frac{1}{2}\frac{h_4 '^2}{h_4^2}.
\end{equation}
\label{ricci_einstein_eq_mu2}
\end{subequations}

\section{One exact solution with back reaction}
\label{sec:mu2_one_exact}

In this short Appendix we would like to report a curious byproduct of our study.
Namely, we found one exact solution of the non-linear Einstein equations \eqref{einstein_eq_mu2_full}.

In order to derive that solution, let us make a substitution as in eq.~\eqref{warp_factors_small-r_power}, only now we will view it as an exact ansatz rather than just the leading-order approximation:
\begin{equation}
	 h_i = C_i  r^{\alpha_i}
\label{warp_factors_power_exact}
\end{equation}
The index runs $i=4,6,a,\omega$ for the warp factors $h_4$, $h_6$, $h_a \equiv a$ and $h_\omega \equiv \omega$ respectively, and $C_i$, $\alpha_i$ are constants. 
After this substitution we obtain a system of equations identical to eq.~\eqref{powers_eq_main_mu2}.
Now, to match the powers of $r$ in all the terms in this equations, we consider the following case:
\begin{equation}
	\alpha_a = 0 \,, \quad
	\alpha_\omega = 0 \,, \quad
	-4-\alpha _4-\alpha _6-\alpha _{\omega } = 0
\end{equation}
In this case, eq.~\eqref{powers_eq_main_mu2} turns into a system of purely algebraic equations.
It has one non-trivial solution with positive warp factors:
\begin{subequations}
\begin{equation}
	C_a = \frac{\alpha_{4}^2}{3} \,,  \quad 
	C_\omega = \frac{3}{2} \,, \quad
	\frac{\mu_2^2}{C_{4} C_{6} g_s} = \frac{1}{4} \,,
\end{equation}
\begin{equation}
	\alpha_{6} = -4-\alpha_{4} \,, \quad
	\alpha_a = 0 \,, \quad
	\alpha_\omega = 0
\end{equation}
\end{subequations}
Therefore, we can write down the warp factors of the metric \eqref{warped_singular_coni_full_metric} as
\begin{equation}
	h_4 = C \left( \frac{r}{\sqrt{\mu_2}} \right)^\alpha \,, \quad
	h_6 = \frac{4}{C g_s} \left( \frac{r}{\sqrt{\mu_2}} \right)^{-4-\alpha} \,, \quad
	a = \frac{\alpha^2}{3} \,, \quad
	\omega = \frac{3}{2}
\label{mu2_one_exact_solution}
\end{equation}
where $\alpha \in \mathbb{R} \setminus \{0\} $ and $C > 0$ are free parameters.
Note that this solution transforms correctly under the reparametrization in eq.~\eqref{reparametrization_transformation}.
We have checked explicitly that the warp factors in eq.~\eqref{mu2_one_exact_solution} indeed solve the initial Einstein equations \eqref{einstein_eq_mu2_full} exactly.

Since the solution \eqref{mu2_one_exact_solution} does not match the large-$r$ asymptotic eq.~\eqref{fluxed_mu2_large-r_solution}, it is not directly relevant for the subject of this paper.
However, it is still an interesting question whether this particular solution has a physical meaning.

One can repeat this exercise for the $\mu_1$ system \eqref{ricci_einstein_eq_mu1}, but in this case we did not find any exact solutions with all the warp factors positive.



\newpage
\addcontentsline{toc}{section}{References}
\begin{thebibliography}{99}






\bibitem{HT1}
A.~Hanany and D.~Tong,
{\em Vortices, instantons and branes,}
JHEP {\bf 0307}, 037 (2003).
[hep-th/0306150].

\bibitem{ABEKY}
R.~Auzzi, S.~Bolognesi, J.~Evslin, K.~Konishi and A.~Yung,
{\em Non-Abelian superconductors: Vortices and
 confinement in ${\mathcal N}=2$  SQCD,}
Nucl.\ Phys.\ B {\bf 673}, 187 (2003).
[hep-th/0307287].

\bibitem{SYmon}
M.~Shifman and A.~Yung,
{\em Non-Abelian string junctions as confined monopoles,}
Phys.\ Rev.\ D {\bf 70}, 045004 (2004)
[hep-th/0403149].

 \bibitem{HT2}
A. Hanany and D. Tong,
{\em Vortex strings and four-dimensional gauge dynamics,}
JHEP {\bf 0404}, 066 (2004)
[hep-th/0403158].

\bibitem{ANO}
A.~Abrikosov, Sov.~Phys. JETP {\bf32}, 1442  (1957);
H.~Nielsen and P.~Olesen, Nucl.~Phys. {\bf B61}, 45 (1973).
[Reprinted in {\em Solitons and Particles}, Eds. C. Rebbi and G. Soliani
(World Scientific, Singapore, 1984), p. 365].

\bibitem{Trev}
D.~Tong, {\em TASI Lectures on Solitons,}
  arXiv:hep-th/0509216.

\bibitem{Jrev}
  M.~Eto, Y.~Isozumi, M.~Nitta, K.~Ohashi and N.~Sakai,
{\em Solitons in the Higgs phase: The moduli matrix approach,}
  J.\ Phys.\ A  {\bf 39}, R315 (2006)
  [arXiv:hep-th/0602170].
  
  \bibitem{SYrev}
M. Shifman and A. Yung,
{\em Supersymmetric Solitons and How They Help Us Understand Non-Abelian Gauge Theories},
  Rev.\ Mod.\ Phys.\  {\bf 79}, 1139 (2007)
  [hep-th/0703267]; for an expanded version see
{\sl Supersymmetric Solitons,}
(Cambridge University Press, 2009).


\bibitem{Trev2}
D.~Tong,
{\em Quantum Vortex Strings: A Review,}
  Annals Phys.\  {\bf 324}, 30 (2009)
  [arXiv:0809.5060 [hep-th]].
	
	
  \bibitem{SYcstring}
 M.~Shifman and A.~Yung,
{\em Critical String from Non-Abelian Vortex in Four Dimensions,}
Phys. \ Lett. {\bf B 750}, 416 (2015)
 [arXiv:1502.00683 [hep-th]].
  
    
    \bibitem{Candel}
P.~Candelas and X.~C.~ de la Ossa,
{\em Comments on conifolds,}
Nucl. \ Phys. \ {\bf B342}, 246 (1990).

\bibitem{NVafa}
A.~Neitzke and  C.~Vafa,
{\em Topological strings and their physical applications},
arXiv:hep-th/0410178.
	
 \bibitem{KSYconifold} 
P. Koroteev, M. Shifman and A. Yung,  
 {\em Non-Abelian Vortex in Four
Dimensions as a Critical String on a Conifold}, 
 Phys. Rev. D 94 (2016)
no.6, 065002 [arXiv:1605.08433 [hep-th]].
	
\bibitem{KSYcstring}
P.~Koroteev, M.~Shifman and A.~Yung,
{\em Studying Critical  String Emerging from 
Non-Abelian Vortex in Four Dimensions},
Phys. \ Lett. \ {\bf B759}, 154 (2016)  
[arXiv:1605.01472 [hep-th]].


	
	\bibitem{SYlittles} 
  M.~Shifman and A.~Yung,
{\em Critical Non-Abelian Vortex in Four Dimensions and Little String Theory,}
  Phys.\ Rev.\ D {\bf 96}, no. 4, 046009 (2017)
  [arXiv:1704.00825 [hep-th]].
	
	

	
	\bibitem{SYlittmult} 
 M.~Shifman and A.~Yung,
 {\em Hadrons of $\mathcal N=2$ Supersymmetric QCD in Four Dimensions from Little String Theory,}
  Phys.\ Rev.\ D {\bf 98}, no. 8, 085013 (2018)
  [arXiv:1805.10989 [hep-th]].
	
	\bibitem{Kutasov}
D.~Kutasov,
{\em Introduction to Little String Theory}, published in
{\sl Superstrings and Related Matters 2001},  Proc. of the ICTP Spring School of Physics, Eds. C. Bachas, K.S. Narain, and   S. Randjbar-Daemi, 2002,  pp.165-209.
	
\bibitem{Y_NSflux}
A.~Yung,
{\em Flux compactification for the critical non-Abelian vortex and quark masses,}
 Phys.\ Rev.\ D {\bf 104},  025007 (2021)
[arXiv:2105.02645 [hep-th]].

\bibitem{NS} 
A. Yung,  
{\em NS Three-form Flux Deformation for the
Critical Non-Abelian Vortex String}, 
Phys. \ Rev. \ D {\bf 106}, 106019 (2022)
[arXiv:2209.08118 [hep-th]].

\bibitem{Louis}
J.~Louis,
{\em Generalized Calabi-Yau compactifications with D-branes and fluxes,}
  Fortsch.\ Phys.\  {\bf 53}, 770 (2005).

  
	
	
\bibitem{Louis2}
J.~Louis and A.~Micu, 
{\em Type II theories compactified on Calabi-Yau threefolds in
the presence of background fluxes,} 
Nucl. \ Phys. \ {\bf B635},  395 (2002) 
[hep-th/0202168].

\bibitem{Kachru}
S.~Kachru, A.~Kashani-Poor,
{\em Moduli Potentials in Type IIA Compactifications with RR and NS Flux}
	JHEP {\em 0503},  066 (2005)
		[arXiv:hep-th/0411279]
		
\bibitem{FI}
  P.~Fayet and J.~Iliopoulos,
{\em Spontaneously Broken Supergauge Symmetries and Goldstone Spinors,}
  Phys.\ Lett.\  B {\bf 51}, 461 (1974).
	
	
\bibitem{T}
D.~Tong,
{\em Monopoles in the Higgs phase,}
Phys.\ Rev.\ D {\bf 69}, 065003 (2004)
[hep-th/0307302].

\bibitem{ISY_b_baryon} 
 E.~Ievlev, M.~Shifman and A.~Yung,
 {\em String baryon in four-dimensional
N=2 supersymmetric QCD from the 2D-4D correspondence,}
  Phys.\ Rev.\ D {\bf 102}, 054026 (2020)
  [arXiv:2006.12054 [hep-th]].

\bibitem{AchVas}
For a review see e.g. A.~Achucarro and T.~Vachaspati,
{\em Semilocal and electroweak strings,}
  Phys.\ Rept.\  {\bf 327}, 347 (2000)
  [hep-ph/9904229].
  
\bibitem{SYsem}
 M.~Shifman and A.~Yung,
  {\em Non-Abelian semilocal strings in  ${\mathcal N} = 2$ supersymmetric QCD,}
  Phys.\ Rev.\  D {\bf 73}, 125012 (2006)
  [arXiv:hep-th/0603134].
  
 \bibitem{Jsem}
M.~Eto, J.~Evslin, K.~Konishi, G.~Marmorini, et al.,
{\em On the moduli space of semilocal strings and lumps,}
  Phys.\ Rev.\  D {\bf 76}, 105002 (2007)
  [arXiv:0704.2218 [hep-th]].
		
		
	\bibitem{SVY}
 M.~Shifman, W.~Vinci and A.~Yung,
{\em Effective World-Sheet Theory for Non-Abelian Semilocal 
Strings in ${\mathcal N} = 2$ Supersymmetric QCD,}
  Phys.\ Rev.\ D {\bf 83}, 125017 (2011)
  [arXiv:1104.2077 [hep-th]].
	
\bibitem{W93}
E.~Witten,
{\em Phases of N = 2 theories in two dimensions,}
  Nucl.\ Phys.\ B {\bf 403}, 159 (1993)
  [hep-th/9301042].

	\bibitem{Ohta}
K.~Ohta and T.~Yokono,
{\em Deformation of Conifold and Intersecting Branes,}
 JHEP {\bf 0002}, 023 (2000)
  [hep-th/9912266].

    \bibitem{KlebStrass}
I.~R.~Klebanov and  M.~J.~Strassler,
{\em Supergravity and a Confining Gauge Theory: Duality Cascades and $chi$SB-Resolution of 
Naked Singularities,}
JHEP {\bf 0008}, 052 (2000)
  [hep-th/0007191].

  


\bibitem{Strom}
A. Strominger,
{\em Massless black holes and conifolds in string theory,}
Nucl. \ Phys. \ B {\bf 451}, 96 (1995)
 hep-th/9504090.



\bibitem{KlebNekras}
I.~R.~Klebanov and N.~A.~Nekrasov, 
{\em Gravity duals of fractional branes and logarithmic
RG flow,}
 Nucl.\  Phys. \ {\bf B574},  263 (2000)
[hep-th/9911096]  


\bibitem{KlebTseytlin}
I.~R.~Klebanov and A.~A.~Tseytlin, 
{\em Gravity duals of supersymmetric SU(N)×SU(N+
M) gauge theories,}
 Nucl. \  Phys. \ {\bf B578},  123 (2000)
[hep-th/0002159]

\bibitem{tseytlin}
	L.~A. Pando~Zayas and A.~A. Tseytlin, 
	{\em 3-Branes on Resolved Conifold},
	Phys. \ Rev. \ D {\bf 63}, 086006 (2001)
	[hep-th/0101043].
 
\bibitem{GiddKachruPolch}
S. B.~Giddings, S.~Kachru and J.~Polchinski,
{\em Hierarchies from Fluxes in String Compactifications,}
 Phys. \ Rev. \ D {\bf 66},  106006 (2002)(2002)
  hep-th/0105097 


	


\bibitem{git_conifold_einstein}
E.~Ievlev, P.~Pichugina,
``{\fontfamily{pcr}\selectfont
conifold-einstein}'',
\url{https://github.com/gvenllian/conifold-einstein}, 2024

 
\bibitem{Johnson:1999qt}
C.~V.~Johnson, A.~W.~Peet and J.~Polchinski,
{\em Gauge theory and the excision of repulson singularities,}
Phys. Rev. D \textbf{61}, 086001 (2000)
[arXiv:hep-th/9911161 [hep-th]].
 	

 
\bibitem{AharSeib}
O.~Aharony, S.~S.~Razamat, N.~Seiberg and B.~Willett, 
{\em The long flow to freedom,}
JHEP {\bf 1702}, 056 (2017) 
[arXiv:1611.02763 [hep-th]]


\end{thebibliography}
\end{document}